\documentclass[12pt,preprint]{aastex}

\usepackage{graphicx,color}

\newcommand{\etal  }{{et al.} }
\newcommand{\msun}{\thinspace M_\odot}  
  
\newcommand{\vect}[1]{\mbox{\boldmath$#1$}}

\newcommand{\rhoc}{\rho_{\rm cri}}
\newcommand{\rhoz}{\rho_0}

\newcommand{\cm  }{\,{\rm cm}^{-3} }
\newcommand{\gcm  }{\,{\rm g\,cm}^{-3} } 

\newcommand{\dfrac}[2]{{\displaystyle \frac{#1}{#2}}  }

\newcommand{\rh  }{r_{\rm H}} 
\newcommand{\rht }{\tilde{r}_{\rm H}} 
\newcommand{\rp  }{r_{\rm p}} 
\newcommand{\ro  }{a_{\rm p}} 
\newcommand{\rs  }{\tilde{r}_{\rm sink}} 
\newcommand{\rst  }{\tl{r}_{\rm sink}} 
\newcommand{\me  }{\thinspace M_\oplus } 

\newcommand{\hs}{h}
\newcommand{\mj}{M_{\rm Jup}}
\newcommand{\rj}{r_{\rm Jup}}
\newcommand{\tl}[1]{\tilde{#1}}

\newcommand{\mdot}{ d\tilde{M}_{\rm p}/d\tilde{t} }

\shorttitle{Gas-Giant Planet Formation}
\shortauthors{Machida  \etal 2007}

\begin{document}
\title{Gas Accretion onto a Protoplanet and Formation of a Gas Giant Planet}

\author{Masahiro N. Machida\altaffilmark{1}, Eiichiro Kokubo\altaffilmark{2}, Shu-ichiro Inutsuka\altaffilmark{1}, and Tomoaki Matsumoto\altaffilmark{3}} 

\altaffiltext{1}{Department of Physics, Graduate School of Science, Kyoto University, Sakyo-ku, Kyoto 606-8502, Japan; machidam@scphys.kyoto-u.ac.jp, inutsuka@tap.scphys.kyoto-u.ac.jp}
\altaffiltext{2}{Division of Theoretical Astronomy, National Astronomical Observatory of Japan, Osawa, Mitaka,
Tokyo 181-8588, Japan; kokubo@th.nao.ac.jp}
\altaffiltext{2}{Faculty of Humanity and Environment, Hosei University, Fujimi, Chiyoda-ku, Tokyo 102-8160, Japan; matsu@i.hosei.ac.jp}


\begin{abstract}
We investigate gas accretion onto a protoplanet, by considering the 
 thermal effect of gas in three-dimensional hydrodynamical simulations,
 in which the wide region from a protoplanetary gas disk to a Jovian
 radius planet is resolved using the nested-grid method.   
We estimate the mass accretion rate and growth timescale of gas giant
 planets. 
The mass accretion rate increases with protoplanet mass for 
 $M_{\rm p}<M_{\rm cri}$, while it becomes saturated or decreases for
 $M_{\rm p}>M_{\rm cri}$, where 
 $M_{\rm cri} \equiv 0.036\,\mj(a_{\rm p}/1{\rm AU})^{0.75}$, and 
 $\mj$ and $\ro$ are the Jovian mass and the orbital radius,
 respectively.  
This accretion rate is typically two orders of magnitude smaller than
 that in two-dimensional simulations.  
The growth timescale of a gas giant planet or the timescale of the
 gas accretion onto the protoplanet is about $10^5$ yr, that is two
 orders of magnitude shorter than the growth timescale of the solid core.
The thermal effects barely affect the mass accretion rate because the
 gravitational energy dominates the thermal energy around the
 protoplanet. 
The mass accretion rate obtained in our local simulations agrees
 quantitatively well with those obtained in global simulations with
 coarser spatial resolution. 
The mass accretion rate is mainly determined by the protoplanet
 mass and the property of the protoplanetary disk.
We find that the mass accretion rate is correctly calculated when the
 Hill or Bondi radius is sufficiently resolved. 
Using the oligarchic growth of protoplanets, we discuss the formation
 timescale of gas  giant planets.
\end{abstract}

\section{INTRODUCTION}
\label{sec:intro}
Currently, more than 400 exoplanets have been observed. Most such planets are considered to be gas giant planets.
Although giant planets are preferentially observed, observations imply that gas giant planets, such as Jupiter and Saturn in our solar system, can be born around stars.
Thus, it is important to understand the formation process of the gas giant planet.
In general, gas planets are formed in the protoplanetary disk (or the circumstellar disk) around the protostar.
However, there is a problems about the growth (or the gas accretion rate) of the gas planet, which is related to the resulting mass of a gas giant planet.

According to the core accretion scenario \citep{perri74, mizuno78, hayashi85}, the protoplanet has a less massive hydrostatic gas envelope when a protoplanet core mass is less than $M_{\rm core} \lesssim 10\me$ where $\me$ is the Earth mass, while the protoplanet captures a massive gas envelope from the protoplanetary disk, that is, a runaway gas accretion phase, to become a gas giant planet when $M_{\rm core} \gtrsim 10\me$ {\citep[e.g.,][]{mizuno78,mizuno80,stevenson82,bodenheimer86,pollack96,ikoma00,ikoma01,hubickyj05}.
Since the gas giant planet acquires almost all of its mass in the runaway accretion phase, the gas flow and mass accretion rate in this phase are important to determine the protoplanet evolution and its resulting mass.
Since gas accretion onto the protoplanet may be closely related to the formation of the circumplanetary disk and acquisition process of angular momentum \citep{machida08,machida09}, three-dimensional calculations are required for investigating the runaway accretion phase.
\cite{machida09} showed that the circumplanetary disk is formed in a compact region near the protoplanet.
Thus,  we may have to resolve the present size of the gas giant planet to estimate the gas accretion rate onto the protoplanet.
As well, to properly handle the outer boundary of the protoplanetary system (protoplanet and circumplanetary disk), the region sufficiently far from the gravitational sphere of the protoplanet, that is, the Hill radius, should also be included, since gas flows into the protoplanetary system from outside the Hill sphere.
Therefore, there is a need to incorporate vastly different spatial scales, ranging from the planet current radius to the Hill radius.
For example, the Hill radius of Jupiter ($r_{\rm H, Jup}=5.4\times10^{12}$\,cm) is about 700 times larger than Jupiter's current radius ($7.1\times10^9$\,cm).

So far, the gas accretion process onto the proto-gas giant planet was investigated mainly in global three-dimensional calculations \citep[e.g.,][]{kley01,dangelo02,dangelo03,bate03,klahr06,dobbs07,paardekooper08,fouchet08,dangelo08}.
However, the fine structures in the proximity of the protoplanet cannot be resolved in such calculations.
A compact disk is formed in the range of $r<10-50\,\rp$ \citep{machida08,machida09}.
Thus, the mass accretion rate may have to be derived by resolving the spatial resolution with  at least $\sim10\,\rp$.
It should be noted that the mass accretion rate obtained from a global simulation can be correct, if the mass accretion rate is determined solely by the global structure around the Hill sphere and the circumplanetary disk contributes little to gas accretion. 
Even in such a case, the mass accretion rate should be investigated in calculations with higher spatial resolution to confirm the validity of the accretion rate derived in simulations with coarser spatial resolution.
There are only a few studies that report the mass accretion rate onto the protoplanet with sufficiently high-spatial resolution in a local simulation \citep{tanigawa02a,ayliffe09}.
However, in these studies, the gas flow from the region outside the Hill radius to the proximity of the protoplanet was not sufficiently discussed. 
Furthermore, the acquisition process of the angular momentum and formation of the circumplanetary disk are not discussed in these studies.

In this paper, we focus on the mass accretion onto the protoplanet in runaway gas accretion phase, in which the gas continues to collapse onto the protoplanet without additional heating by collision of planetesimals.
After the solid core (or planetary core) formation, the formation process of the gas giant planet can be divided into two phases.
The quasi-static envelope slowly contracts with a timescale of $\sim 10^6-10^7\,$yr \citep[e.g.,][]{ikoma00} when the protoplanet is less massive than $M_{\rm core} \lesssim 1-10\me$,  while the gas rapidly collapses onto the protoplanet when $M_{\rm core} \gtrsim 1-10\me$.
We investigate the evolution of the protoplanetary system only in the runaway gas accretion phase, based on the results of a local simulation with a sufficiently high-spatial resolution using the nested-grid method, in which the region of $\sim5-20\,\rh$ from the protoplanet is resolved with cells having the size of the radius of the present-day Jupiter.
Since the acquisition process of the angular momentum \citep{machida08} and circumplanetary disk formation \citep{machida09} around the protoplanet were already investigated using this method, this paper will focus on gas accretion onto the protoplanet and gas flow in the proximity of the protoplanet.
As a result of calculation, we found that the mass accretion rate is correctly calculated when the Hill or Bondi radius is sufficiently resolved, and it barely depends on the thermal effect around proto-gas giant planet.
The structure of the paper is as follows. 
\S 2 gives the model frameworks, while \S 3 describes the numerical methods used. 
The numerical results are presented in \S 4 and compared with the results of previous studies in \S 5.
\S 6 discusses protoplanetary growth. The conclusions of this paper are presented in \S 7.

\section{MODEL}
\subsection{Basic Equations}
\label{sec:basic}
A local region around a protoplanet is considered using the shearing sheet model \citep[e.g.,][]{goldreich65}, in which the self-gravity of the protoplanetary disk is ignored.
In addition, no physical viscosity is included, and the numerical viscosity can be ignored because it is sufficiently small.
Thus, an inviscid gas disk model is adopted.
The orbit of the protoplanet is assumed to be circular in the equatorial plane of the protoplanetary  disk. 
Local rotating Cartesian coordinates with the origin at the protoplanet are set up, in which the $x$-, $y$-, and $z$-axis are the radial, azimuthal, and vertical directions of the disk.  
The equations of hydrodynamics without self-gravity are solved [see, eqs. (1)-(6) of \citealt{machida08}].

For the gas, a barotropic equation of state is adopted (for details, see \citealt{machida09}).
In a local region, the protoplanetary disk has an almost constant temperature \citep{hayashi85}, while the gas around the protoplanet, that is, the gas envelope, has a higher temperature than the protoplanetary disk \citep{mizuno78,mizuno80,bodenheimer86,pollack96,ikoma00}.
\citet{mizuno78} studied the structure and stability of the envelope around the protoplanet, on the assumption that the envelope is spherically symmetric and in hydrostatic equilibrium.
They also investigated the thermal evolution of the envelope, parameterizing the dust grain opacity, and determined the boundary between the isothermal and adiabatic regions.
Using Figure~2 of \citet{mizuno78}, the thermal evolution around the protoplanet is modeled as a function of the gas density, that is, using the barotropic equation of state, as 
\begin{equation}
 P = c_{\rm s,0}^2 \rho \, \left[1-{\rm tanh}\left(\dfrac{\rho}{\rho_{\rm cri}} \right) \right]  
+ K \rho^\gamma \, {\rm tanh}\left(\dfrac{\rho}{\rho_{\rm cri}} \right),
\label{eq:eos} 
\end{equation}
 where $c_{\rm s}$ is the sound speed, $\gamma$ is the adiabatic index ($\gamma=1.4$),  and 
the adiabatic constant $K$ is defined as 
$
K = c_{\rm s,0}^2 \rho_{\rm cri}^{1-\gamma},
$
where $\rho_{\rm cri}$ is  the critical density, wherein the gas behaves isothermally for  $\rho < \rho_{\rm cri}$, and adiabatically for $\rho > \rho_{\rm cri}$.
 In this study, $\rho_{\rm cri}$ = $\infty$ (isothermal model), $\rhoz$, 10$\rhoz$, $10^2\,\rhoz$, and $10^3\,\rhoz$ (adiabatic model) are used, where $\rhoz$ is the initial density on the equatorial plane.   
The hyperbolic tangent (tanh) function is used to smoothly connect the first (isothermal) and second (adiabatic) terms in Equation~(\ref{eq:eos}).
The thermal evolution for different $\rho_{\rm cri}$ is plotted against the gas density in Figure~\ref{fig:1}, in which the gas temperature is constant in the isothermal model ($\rho_{\rm cri}= \infty$), while it increases gradually from the initial value at $\rho\sim \rho_{\rm cri}$  in the adiabatic models ($\rho_{\rm cri}$ = $\rhoz$, 10$\rhoz$, $10^2\,\rhoz$, and $10^3\,\rhoz$).

In the standard disk model \citep{hayashi85}, at a Jovian orbit, the density is $\rhoz =1.5\times 10^{-11}\gcm$, and the temperature is $T_0=123$\,K. 
Thus, for example, in a model with $\rhoc = 10\,\rhoz$, the gas behaves isothermally when $\rho \ll 1.5\times10^{-10}\gcm$, while it behaves adiabatically when $\rho \gg 1.5\times 10^{-10}\gcm$.
Comparing Figure~\ref{fig:1} with Figure~2 of \citet{mizuno78}, the thermal evolution of the model with $\rho_{\rm cri}=10\, \rhoz$  (Fig.~\ref{fig:1} broken line) corresponds to that for a gas envelope around a proto-Jovian planet (Fig.2 of \citealt{mizuno78}) with a dust opacity of $\kappa_{\rm g}=1.0\times 10^{-2}$\,cm$^2$\,g$^{-1}$.
In this model setting, the critical density $\rho_{\rm cri}$ corresponds to the dust opacity $\kappa_{\rm g}$ in \citet{mizuno78}.
\citet{mizuno78} adopted $\kappa_{\rm g}=1.0\times 10^{-4}$\,cm$^2$\,g$^{-1}$ as the most reliable parameter of a proto-Jovian planet, indicating that a more realistic gas temperature of the envelope is lower than that in the model with $\rho_{\rm cri} = 10\,\rhoz$ (the dotted line of Figure~\ref{fig:1}).
In our settings, because $\kappa_{\rm g}=1.0\times 10^{-4}$\,cm$^2$\,g$^{-1}$ almost corresponds to $\rhoc=10^2\rhoz$, we call the models having $\rhoc = 10^2\rhoz$ `most realistic models'.
It should be noted that the critical density $\rho_{\rm cri}$ increases as the dust opacity $\kappa_{\rm g}$ decreases.
Thus, in models with $\rhoc=10\rhoz$, the thermal energy around the protoplanet may be overestimated.
On the other hand, when the isothermal equation of state is adopted, the thermal energy around the protoplanet is obviously underestimated.
Therefore, it is expected that the actual thermal evolution is located between models with $\rhoc=\rhoz$ and $\rho_c=\infty$.

\subsection{Protoplanetary Disk Model and Scaling}
The initial settings are similar to \citet{miyoshi99}, \citet{machida06b}, \citet{machida08}, and \citet{machida09}.
The gas flow has a constant shear in the $x$-direction as
\begin{equation}
\vect{v_0} = ( 0,\,  -{3}/{2}\,\Omega_{\rm p}\, x, \,0 ),
\label{eq:shear}
\end{equation}
 where $\Omega_{\rm p}$ is the Keplerian angular velocity of the protoplanet
\begin{equation}
\Omega_{\rm p} = \left( \dfrac{ G\, M_{\rm c} } { \ro^3} \right)^{1/2},
\label{eq:omegap}
\end{equation}
 where $G$ is the gravitational constant, $M_{\rm c}$ is the mass of the central star, and $\ro$ is the orbital radius of the protoplanet.

For hydrostatic equilibrium, the density is given by 
$
\rho_0 = {\sigma_0}\, {\rm exp } \left(- {z^2}/{2 h^2} \right) /{\sqrt{2\pi}h} 
, 
$
 where $\sigma_0$ ($\equiv \int_{-\infty}^{\infty} \rho_0 \, dz $) is the
 surface density of the unperturbed disk. 
The scale height $h$ is related to the sound speed $c_{\rm s}$ by
 $h=c_{\rm s}/\Omega_{\rm p}$. 

Using the standard solar nebular model \citep{hayashi81,hayashi85}, the temperature $T$, sound speed $c_{\rm s}$,  and gas density $\rho_{c, 0}$ can be described as
\begin{equation}
T = 280 \left( \dfrac{L}{L_{\odot}} \right)^{1/4}
 \left(\dfrac{\ro}{1\,{\rm AU}} \right)^{-1/2}\ \ {\rm K},
\label{eq:nebular-temp}
\end{equation}
 where $L$ and $L_{\odot}$ are the protostellar and solar luminosities,  
\begin{equation}
 c_{\rm s} = \left( \dfrac{k\,T}{\mu m_{\rm H}} \right)^{1/2} =
 1.9\times 10^4\, \left( \dfrac{T}{10\,{\rm K}} \right)^{1/2} \, \left( \dfrac{2.34}{\mu} \right)^{1/2} \ \  {\rm cm\,s^{-1}}, 
\label{eq:nebular-cs}
\end{equation}
 where $\mu =2.34$ is the mean molecular weight of the gas composed mainly of H$_2$ and He, and   
\begin{equation}
\rhoz = 1.4 \times 10^{-9} \left( \dfrac{\ro}{1\,{\rm AU}} \right)^{-11/4} \ \ {\rm g}\cm.
\label{eq:nebular-dens}
\end{equation}
When $M_c = 1\msun$ and $L=1\,L_{\odot}$ are adopted, the scale height $h$ can be described as
\begin{equation}
h = 5.0\times 10^{11} \left( \dfrac{\ro}{1 {\rm AU}} \right)^{5/4}  \ \ \ {\rm cm}.
\label{eq:negular-scale-height}
\end{equation}
The inverse of the angular velocity $\Omega_{\rm p}$  is described as
\begin{equation}
\Omega_{\rm p}^{-1} = 0.16 \left( \dfrac{\ro}{1 {\rm AU}} \right)^{3/2} \ \ \ {\rm yr}.
\label{eq:omegat}
\end{equation}

The basic equations can be normalized using unit time, $\Omega_{\rm p}^{-1}$, and unit length, $\hs$.
The density is also scalable and is normalized using $\sigma_0/h$.
Hereafter, the normalized quantities are expressed with a tilde on top, for example, $\tilde{x} = x/h$, $\tilde{\rho} = \rho_0/(\sigma_0/h)$, and $\tilde{t}=t\,\Omega_{\rm p}$. Further details can be found in \citet{machida08}. 
A dimensionless description is given by Equations~(14)-(19) of \citet{machida09}.
The dimensionless quantities are converted into dimensional quantities using Equations~(\ref{eq:nebular-temp})--(\ref{eq:omegat}).
The gas flow is characterized by two parameters, the dimensionless Hill radius $\rht=\rh/h$, and the critical density $\tilde{\rho}_{\rm cri}$.
The dimensional Hill radius is defined by
\begin{equation}
 r_{\rm H} = \left( \dfrac{M_{\rm p}}{3M_{\rm c}} \right)^{1/3} \ro.
\label{eq:hill}
\end{equation} 
In this study, a Hill radius $\rht$ ranging from 0.29 to 1.36 is used.
As a function of the orbital radius and the mass of the central star, the parameter $\rht$ is related to the actual protoplanet mass in units of Jovian mass $\mj$ by 
\begin{equation}
 \dfrac{M_{\rm p}}{\mj} = 
 0.12 \left( \dfrac{M_{\rm c}}{1\msun} \right)^{-1/2} 
 \left( \dfrac{\ro}{1\,{\rm AU}} \right)^{3/4} \, \rht^3.
\label{eq:mass-to-hill}
\end{equation}
For example, in the model with $\rht = 1.0$,  $\ro=5.2$\,AU and
 $M_{\rm c} = 1\,\msun$, the protoplanet mass is 
 $M_{\rm p} = 0.4 M_{\rm J}$.
In the parameter range of $\rht =0.29-1.36 $, at a Jovian orbit ($\ro=5.2$\,AU), protoplanets have masses of $0.01\mj-1\mj$.
Table 1 gives the dimensionless Hill ($\rht$) and Bondi ($\tl{r}_{\rm B}$) radii, the masses of protoplanets at Jovian (5.2\,AU) orbit, the critical densities and sink radii for all models.
Model names consist of two parts: the protoplanet mass at the Jovian orbit and the critical density.
For example, model M001A3 has parameter values given as $M_{\rm p}=0.01\mj$, $\rho_{\rm cri}= 10^3\,\rhoz$.

\section{NUMERICAL METHODS}
\subsection{Numerical Procedures}
\label{sec:procedure}
The purpose of this study is to investigate gas accretion onto the protoplanet  in three-dimensional simulations.
However, given current CPU limitations, it is impossible to calculate the complete evolution of the gas giant planet with sufficiently high-spatial resolution, that is, the evolution of the planet from a solid core ($\sim10\me$) with a thin gas envelope to a protoplanet that acquires a massive atmosphere ($\sim1\mj$) cannot be performed.
Thus, the mass accretion rate is derived using the following procedure:
(i) With a fixed protoplanet mass, the evolution of the protoplanetary system until the gas flow reaches the steady state ($\sim100$\,orbits) is calculated,
(ii) Then a  sink is introduced, and the mass accretion rate that is derived is averaged over a further $\sim20$\,orbits, and 
(iii) Finally, steps (i) and (ii) are repeated by changing the protoplanet mass to give the growth rate of the protoplanet.

\subsection{Nested-Grid Method}
\label{sec:nested-grid}
To investigate the formation of a circumplanetary disk in a protoplanetary disk, it is necessary to cover a large dynamic spatial scale range.
Using the nested-grid method \citep[for details, see][]{machida05, machida06a}, the regions near and remote from the protoplanet are covered with adequate resolution. 
Each level of a rectangular grid has the same number of cells ($ = 64 \times 128 \times 16 $), but cell width $\Delta \tilde{s}(l)$ depends on the grid level $l$. 
The cell width is reduced by 1/2 with increasing grid level ($l \rightarrow l+1$).
Eight grid levels ($l_{\rm max}=8$) are used.
The box size of the  coarsest grid, $l=1$, is $(\tl{L}_x, \tl{L}_y, \tl{L}_z) = (12, 24, 3)$, and that of the finest grid, $l=8$, is $ (\tl{L}_x, \tl{L}_y, \tl{L}_z) = (0.09375, 0.1875, 0.0234)$. 
The cell width in the coarsest grid, $l=1$, is $\Delta \tilde{s} = 0.1875$, and it decreases with $\Delta \tilde{s}=0.1875/2^{l-1}$ as the grid level $l$ increases.
Thus, the finest grid has  $\Delta \tilde{s}(8)\simeq 1.46\times10^{-3}$.
The fixed boundary condition in the $\tilde{x}$- and $\tilde{z}$-direction, and the periodic boundary condition in the $\tilde{y}$-direction are used.

\subsection{Sink Cell and Smoothing Length}
\label{sec:sink}
In the finest grid ($l_{\rm max}=8$), the cell width is $\Delta \tl{s} = 1.46\times 10^{-3}$.
In real units, when the protoplanet is located at 5.2\,AU, the cell width corresponds to $\Delta s = 5.7\times 10^{9}$\,cm, or 0.8 times the Jovian radius.
The evolution of the protoplanetary system is calculated using a sink.
In the fiducial models, the radius of the sink is $\tilde{r}_{\rm sink}=3.53\times 10^{-3}$ or twice the Jovian radius at Jovian orbit (for details, see~\S\ref{sec:convergence}).
During the calculation, the gas from the region inside the sink radius is removed in each time step.

The smoothing length for the gravitational potential of the protoplanet is not explicitly used.
For numerical calculations, the physical quantities are defined at the cell centre, while the origin (protoplanet's position) is defined as the cell boundary.
Thus, the region inside $ \tilde{r} < \tilde{r}_s \equiv  \sqrt{3} \Delta \tilde{s}(l_{\rm max})/2 $ has a uniform gravitational potential.
At a Jovian orbit, $r_s$ is $0.7$ times the Jovian radius ($\tilde{r}_s = 1.26 \times 10^{-3}$ or $r_s = 4.9\times10^9$\,cm).

\subsection{Convergence Test for the Accretion Rate}
\label{sec:convergence}
To check the convergence of the mass accretion rate onto the protoplanet, the evolution of the protoplanetary system for different sink radii was calculated.
The convergence of other quantities that change with the cell width or grid level with and without a  sink were already investigated in \citet{machida08} and \citet{machida09}.
Figure~\ref{fig:2} shows the mass accretion rate as a function of the sink radius for models M02ISS, M02ISM, M02I, and M02ISL.
The mass accretion rate is derived based on the procedure outlined in \S\ref{sec:procedure}.
As listed in Table~\ref{table:1}, these models have the same Hill radius $\rht$ (or the same protoplanetary mass) and different sink radii $\rst$.
Models have $M_{\rm p} = 0.2\mj$ at Jovian orbit ($\ro=5.2$\,AU), and the isothermal equation of state is used.
Model M02ISL has the smallest sink radius of $\tl{r}_{\rm sink} = 1.6\times 10^{-3}$ which corresponds to $6.3\times 10^9$cm (0.87\,$\rj$), while model M02ISL has the largest sink radius of $\tl{r}_{\rm sink} = 1.1\times 10^{-2}$ which corresponds to $4.3\times10^{10}$cm (6\,$\rj$).
The protoplanetary system was calculated for $\sim100$\,orbits, for which it was confirmed that the gas flow achieves a steady state in $\tl{t}\ll100$\,orbits.
\citet{tanigawa02a} also showed that steady state is reached after $1-10$\,orbits for a local calculation.
\citet{machida09} showed that the gas envelope and angular momentum around the protoplanet reach steady state after 1-10\,orbits.

Figure~\ref{fig:2} shows that the models M02ISS, M02ISM, and M02I have almost the same accretion rate of $\mdot \simeq 0.18$, for models, which have sink radii of $\tl{r}_{\rm sink}<3.5\times10^{-3}$ ($1.4\times10^{10}$cm; 1.92\,$\rj$).
On the other hand, model M02ISL has a sink radius of $\tilde{r}_{\rm sink}=1.1\times10^{-2}$ ($4.3\times10^{10}$cm; 6\,$\rj$), and has a slightly larger accretion rate of $\mdot \simeq 0.35$ than the others.
Thus, the difference in the accretion rate between M02ISS (0.2) and M02ISL (0.35) is only within a factor of 2.
As a result, the accretion rate can safely be estimated when the sink radius is less than $\rs <3.5\times 10^{-3}$.
In the following, we show results  adopting $\rs = 3.5\times 10^{-3}$ ($1.4\times10^{10}$cm; $1.9\,\rj$).

Since the gas inside the sink radius is removed at each time step as mentioned in \S\ref{sec:sink}, the structure, that is, the protoplanet's atmosphere and envelope, inside the sink radius cannot be investigated.
However, it is expected that the gas falling into the sink cannot escape from the protoplanet and, hence, does not influence the results obtained.
In the region where $r<r_{\rm sink}$, since the gravitational energy greatly dominates the thermal energy \citep{mizuno78,machida09}, the gas that falls into the sink cannot be pushed out by the gas pressure.
In addition, the gas falling into the sink does not acquire additional angular momentum inside the sink radius (or near the protoplanet), because the angular momentum of the system is only acquired by the shearing motion of the protoplanetary disk \citep{machida08}.
However, it is unknown how the gas trapped by the gravitational potential of the protoplanet reaches the surface of the protoplanet because the centrifugal force is expected to prevent the gas from further falling inside the sink.
To understand the surface and structure of the protoplanet, the region inside the protoplanet, that is,  inside $r<r_{\rm sink}$ needs to be resolved. 
Such simulation would require an enormous amount of CPU time.
In the calculations, the sink radius is $r_{\rm sink}=1.9\,\rj$ that is smaller than the Roche limit $a_{\rm R}=2.5\,\rp$.
Thus, it is expected that the gas inside $r< r_{\rm sink}$ loses its angular momentum by tidal interaction and finally falls into the gas planet. 
In addition, \citet{klahr06} suggested that the radius of a young planet is about twice the current size of Jupiter.
Therefore, when the sink radius is comparable to the size of the present planet, it is possible to safely estimate the gas accretion rate.

\section{Gas Flow and Mass Accretion Rate}
\label{sec:results}
\subsection{Spiral Pattern and a Circumplanetary Disk}
Figure~\ref{fig:3} shows the structure around the Hill sphere for each model at $\sim100$ orbits ($\tl{t}\simeq 640$).
The systems in Figure~\ref{fig:3} are in steady state, because the gas flow achieves a steady state in $\sim10$\,orbits ($t\simeq 6.3-63$) as shown in \S\ref{sec:convergence} \citep[see also][]{miyoshi99,tanigawa02a,machida08}. 
Models in the figure have a parameter of $\rhoc=10^2 \rhoz$, for which the gas temperature increases adiabatically for $\rho>10^2\rho_{0}$.
The figures show that the global structure in the models is almost the same as for the isothermal models \citep[for details, see][]{machida09}.
In calculations using the isothermal equation of state, the spiral patterns that are distributed from the upper-left to the lower-right region are observed  \citep[e.g.,][]{lubow99}.
The same patterns are seen also in Figure~\ref{fig:3}, in which the barotropic equation of state is used. This can be explained by noting that for regions far from the protoplanet, the gas density is not high enough, and the gas behaves isothermally.
As well, Figure~\ref{fig:3} shows that as the protoplanet mass increases, stronger shock waves appear and the spiral patterns become clearer.
Furthermore, the density contrast between spiral and gap becomes stronger as the protoplanet mass increases.
Thus, it can be concluded that the features observed in adiabatic models correspond well with those features observed in isothermal models.

To focus on the region near the protoplanet, the structures inside the Hill sphere are shown in Figure~\ref{fig:4} in three dimensions.
The grid level of Figure~\ref{fig:4} is $l=6$, while that of Figure~\ref{fig:3} is $l=2$.
Thus, Figure~\ref{fig:4} is a 16 times enlargement of the central part of Figure~\ref{fig:3}.
In Figure~\ref{fig:4}, the region of $\rho>10^{3}\rho_0$ is plotted using the red constant density surfaces, while that for $\rho>10^2\rho_0$ is plotted using the orange constant density surfaces.
It should be noted that the orange and red surfaces do not appear in model M001A2, because this model has a very small part that has $\rho>10^2\rho_0$.
The density distributions on $x=0$, $y=0$, and $z=0$ plane are projected onto each wall surface.
Figure~\ref{fig:4} shows that the red and orange surfaces increase with the protoplanet mass, indicating that a massive protoplanet has a  denser, or more massive, envelope.
The orange surface flattens as the protoplanet mass increases, while the red surface in all models except for model M001A2 has a sufficiently flattened structure.
\citet{machida08} and \citet{machida09} showed that the specific angular momentum of the envelope increases in proportional to $j\propto M_{\rm p}$ when the protoplanet mass is smaller than $M<\mj$.
Thus, the angular momentum also increases with the protoplanet mass.
Therefore, the centrifugal radius increases with the protoplanet mass. Since the centrifugal force gradually affects the more distant region from the protoplanet as the protoplanet mass increases, the low-density region gradually flattens out owing to the centrifugal barrier.
On the other hand, the red surfaces always have flattened structures, because the red surface (or higher density envelope) is located near the protoplanet that is inside the centrifugal radius even when the protoplanet is less massive $M_{\rm p}\simeq0.05\mj$.

\subsection{Mass Accretion Rate onto a Protoplanet}
\label{sec:accretionrate}
Figure~\ref{fig:5} shows the mass accretion rates onto the protoplanet for all models against the cube of the Hill radius, $\rht^3$.
With a fixed protoplanet's orbit, $\rht^3$ can be connected to the protoplanet mass using Equation~(\ref{eq:mass-to-hill}).
As a reference, the mass accretion rate and the protoplanet mass at a Jovian orbit are given in the upper and right axes in Figure~\ref{fig:5}.
The mass accretion rate and the protoplanet mass at the Jovian orbit can be converted into those at any orbit using the value of $\ro$ in parenthesis in the upper and right axes.
In this paper, the mass accretion rate at the Jovian orbit will be used for convenience.

As shown in Figure~\ref{fig:5}, the mass accretion rates increases rapidly for protoplanetary mass in the range of $M_{\rm p}\lesssim 0.2\mj$.
The accretion rates for isothermal and adiabatic models are almost the same in this range.
The accretion rate for each model has a peak at $M_{\rm p}\simeq 0.2\mj$, then it gradually decreases in the range of $M_{\rm p}\gtrsim0.2\mj$.
As well, in this range, the accretion rates depend slightly on the equation of state used.
The difference increases as the protoplanet mass increases.
For $M_{\rm p}\gtrsim 0.2\mj$, the decrease of the mass accretion rate in models with a harder equation of state, that is smaller $\rhoc$, is less than in models with a softer equation of state, that is,  larger $\rhoc$.
For example, the barotropic model of $\rho_{\rm cri}=10\, \rho_{\rm 0}$ has accretion rates of $\dot{M}_p=4.7\times10^{-5}\mj$\,yr$^{-1}$ at $M=0.2\mj$, and $4.1\times10^{-5}\mj$\,yr$^{-1}$ at $1\mj$.
On the other hand, the isothermal model shows a steeper decrease of the accretion rate than the other models.
The isothermal model had  accretion rates of $\dot{M}_p=4.3 \times 10^{-5}\mj$\,yr$^{-1}$ at $M_{\rm p}=0.2\mj$, and  $2.3\times10^{-5}\mj$\,yr$^{-1}$ at $M=1\mj$.
Thus, the difference between isothermal and barotropic models is not so large even when the protoplanets become massive.
When the protoplanet has a Jovian mass $1\mj$, the accretion rate in model with $\rhoc=\rhoz$ is only 2.4 times larger than that for the isothermal model.

As shown in \S\ref{sec:basic},  it is expected that the actual mass accretion rate is located between the isothermal and the adiabatic model with $\rho_{\rm cri} = \rho_0$.
Around the protoplanet, the thermal energy is underestimated in the isothermal model, while it is overestimated in the model with $\rho_{\rm cri} = \rho_0$.
We fitted the mass accretion rate by the solid line in Figure~\ref{fig:5}, in which, for convenience, it is fitted as a constant in the range of $M_{\rm p}>0.125\mj$.
It should be noted that, although the accretion rate in each model gradually decreases in this range, the rate of decrease is sufficiently small especially in adiabatic models.
As well, it can be noted that, since the protoplanetary evolution in the range of $M_{\rm p}>1\mj$ was not calculated, the accretion rate may not apply in the range of $M_{\rm p}>1\mj$.
Using dimensionless quantities, the mass accretion rate is described as
\begin{equation}
\dfrac{d\tl{M}_{\rm p}}{d\tl{t}} \simeq \left\{
\begin{array}{ll}
0.83\, (\rht^3)^{3/2} \   {\rm for}  \ \rht^3 < 0.3 \\
0.14  \ \ \ \  \ \ \ \ \ \ \, {\rm for}  \ \rht^3 > 0.3. 
\end{array}
\right.
\label{eq:mdot1}
\end{equation}
In Equation~(\ref{eq:mdot1}), the mass accretion rate for $\rht^3>0.3$ ($\dot{M}_{\rm p}=0.14$)  almost corresponds to that of the average ($\dot{M}_{\rm p}=0.16$) in most realistic models having $\rho=10^2\rhoz$ for $\rht>0.3$. 
Since the protoplanetary evolution was calculated using a local simulation, it is possible to convert the  mass accretion rate from the fixed orbit to any arbitrary orbit, arbitrary density, and temperature of the protoplanetary disk model.
Using Equations~(\ref{eq:nebular-temp})-(\ref{eq:omegat}), the dimensional mass accretion rate can be described as
\begin{eqnarray}
\dfrac{dM_{\rm p}}{dt} \left[ \dfrac{M_{\rm Jup}}{yr} \right]  \simeq  \hspace{12cm} \nonumber \\
\left\{
\begin{array}{ll}
1.2 \times 10^{-2}\,  \left( \dfrac{M_{\rm p}}{M_{\rm Jup}} \right)^{3/2} \left( \dfrac{\ro}{{\rm 1 AU}} \right)^{-13/8}
 \left( \dfrac{L}{L_\odot}\right)^{-3/16} \left( \dfrac{M_{\rm c}}{\msun}\right)^{-1/4}
\    
\\ \hspace{50mm} {\rm for}  \ M_{\rm p}/\mj < 0.036 \left( \dfrac{\ro}{\rm 1 AU} \right)^{3/4} \left(\dfrac{M_{\rm c}}{\msun} \right)^{-1/2} \left( \dfrac{L}{L_\odot}\right)^{3/8}\\ 
\\
8.1\times 10^{-5}\, \left( \dfrac{\ro}{{\rm 1 AU}} \right)^{-1/2} 
\left( \dfrac{L}{L_\odot}\right)^{3/8} \left( \dfrac{M_{\rm c}}{\msun}\right)^{-1}
\    
\\ \hspace{50mm} {\rm for} \ \ M_{\rm p}/\mj > 0.036 \left( \dfrac{\ro}{\rm 1 AU} \right)^{3/4} \left(\dfrac{M_{\rm c}}{\msun} \right)^{-1/2} \left( \dfrac{L}{L_\odot}\right)^{3/8}.
\end{array}
\right.
\label{eq:mdot}
\end{eqnarray}
When the central protostar has a mass of $M=1\msun$, a luminosity of $L=1L_{\odot}$, and  the standard model \citep[i.e., MMSN disk model; ][]{hayashi85} with surface density $\Sigma_0$ is adopted in Equation~(\ref{eq:mdot}), the accretion rate  at any orbit can be described as
\begin{eqnarray}
\dfrac{dM_{\rm p}}{dt}  \left[ \dfrac{M_{\rm Jup}}{yr} \right] \simeq  \hspace{12cm} \nonumber \\
\left\{
\begin{array}{ll}
1.2 \times 10^{-2}\,  \left( \dfrac{M_{\rm p}}{M_{\rm Jup}} \right)^{3/2} \left( \dfrac{\ro}{{\rm 1 AU}} \right)^{-13/8} \left( \dfrac{\Sigma}{\Sigma_0} \right)    
\hspace{5mm} {\rm for}  \ \left(\dfrac{M_{\rm p}}{\mj} \right)< 0.036 \left( \dfrac{\ro}{\rm 1 AU} \right)^{3/4} 
\\
8.1\times 10^{-5}\, \left( \dfrac{\ro}{{\rm 1 AU}} \right)^{-1/2}  \left( \dfrac{\Sigma}{\Sigma_0} \right)
 \hspace{28mm} {\rm for} \ \left(\dfrac{M_{\rm p}}{\mj} \right) > 0.036 \left( \dfrac{\ro}{\rm 1 AU} \right)^{3/4},
\end{array}
\right.
\label{eq:mdot2}
\end{eqnarray}
where $\Sigma$ is arbitrary surface density at any orbit.
Equation~(\ref{eq:mdot2}) indicates that the growth timescale of the Jovian mass protoplanet at a Jovian orbit is $\sim10^5$\,yr.
We will present a detailed discussion of the mass accretion rate and the growth rate of the protoplanet in \S 5.
The following sections show the gas flows at large (\S\ref{sec:large}) and small scales (\S\ref{sec:small}) around the protoplanet to confirm the relationship between the accretion rate and the circumplanetary disk.

\subsection{Gas inside the Hill Sphere}
\label{sec:large}
This section examines the large-scale gas flows in order to determine gas accretion from outside the Hill sphere. 
Each top panel in Figure~\ref{fig:6} plots the gas streamlines falling into the protoplanet.
Figure shows only models with $\rho_{\rm cri}=10^2\rho_0$, which is the most realistic parameter \citep{mizuno78,machida09}.
There are no differences in the streamlines between models with different equations of state, because the gas outside the Hill sphere, that is, at a large-scale, has a low density and behaves almost isothermally.
In the figure, the streamlines are inversely integrated from the region inside the Hill sphere.
Thus, only the streamlines accreting onto the protoplanetary system are plotted.
Colors of the streamlines indicate the velocity in the vertical direction ($v_z$).
For example, the gas rises upward in blue and green parts of the streamline, while it goes down in yellow and red parts.
The Bondi, which is shown using a black circle,
\begin{equation}
r_{\rm B} = \dfrac{G M_{\rm p}}{c_{s,0}^2},
\label{eq:bondi}
\end{equation}
the doublewide Bondi, shown with a black-dotted circle, the Hill, shown with a white circle, and the doublewide Hill, shown with a white-dotted circle radius, are also plotted in the figure.
Each top panel in Figure~\ref{fig:6} shows that the gas flow falling into the protoplanetary system is controlled either by the Bondi or the Hill radius.
When the Bondi radius is smaller than the Hill radius (models M001A2 and M005A2), the gas flows  into the protoplanet only in the narrow band of $r_{\rm B}\lesssim\vert x \vert \lesssim 3\,r_{\rm B}$.
The bandwidth of streamlines in the region far from the Hill sphere almost corresponds to the Bondi radius ($\sim r_{\rm B}$, see the lower left corner in the top panel of Fig.~\ref{fig:6}).
When the protoplanet mass increases and the Bondi radius exceeds the Hill radius (M04A2, M08A2, and M1A2), the gas flows into the protoplanet in the band of $r_{\rm H}\lesssim \vert x \vert \lesssim 3\,r_{\rm H}$, in which the streamlines in the region far from the Hill sphere have a bandwidth of the Hill radius ($\sim \rh$).
It should be noted that in Figure~\ref{fig:6}, the grid-scale is different in each of the panels.
Thus, it can be concluded that the gas flows into the protoplanetary system occur only in the region given as
\begin{equation}
r_{\rm acc} \lesssim x \lesssim 3\,r_{\rm acc}, \ \ \ \  r_{\rm acc} = {\rm min}(r_{\rm B}, r_{\rm H}).
\end{equation}
It was also shown by \citet{dangelo08} that the mass accretion rate is controlled by either the Bondi or the Hill radius.

The colors of the streamlines imply that the gas rises slightly in regions far from the protoplanet (see, for example, Models M04A2, M08A2, M1A2), then it gradually goes down as the streamlines approach the Bondi or Hill sphere.
Near the Bondi or Hill sphere, the gas rises strongly upward near the shock front and then rapidly falls into the Bondi or Hill spheres in the vertical direction \citep[for details, see][]{machida08}.
The similar features of the gas flow (or streamlines) around the Hill sphere are seen in \citet{kley01}, \citet{dangelo03} and \citet{paardekooper08}.

For each bottom panel in Figure~\ref{fig:6}, the $xz-$plane is plotted, in which the color indicates the mass flux $\rho \vert v \vert$ per unit area.
In the plane, only gas passing through the gray region (or gray points) can reach the protoplanetary system (compare the gray regions in each bottom panel with the streamlines in each top panel). 
Most of the gray points are distributed in the region far from the equatorial plane, indicating that the gas present in the midair flows preferentially into the protoplanetary system, while the gas near the equatorial plane barely flows into it.
Although the gas near the equatorial plane also flows into the Hill sphere, it flows out from the Hill sphere afterwards, as shown in \citet{machida08} and \citet{paardekooper08}.
Figure~\ref{fig:7} shows the streamlines for model M04A2, in which the streamlines are integrated from the area of $\rht<x<2\,\rht$ and $0<z<1.5\,h$ on the $y=-6\,h$ plane, that is, the region denoted by the black squares.
It should be noted that the streamlines in Figure~\ref{fig:6} are {\it inversely} integrated from the region inside the Hill sphere, while, in Figure~\ref{fig:7}, they are integrated along the flow from inside the black square.  
The color of each streamline indicates the mass flux $\rho \vert v \vert$ per unit volume at each mesh point.
The gas flow through the black square in Figure~\ref{fig:7} tracks three different paths: (i) the flow is attracted toward the protoplanet and returns to a similar orbit as the original orbit  (the pass-by region), (ii) the flow is trapped by the protoplanet and falls into the protoplanetary system (the planetary atmosphere region), and (iii) the flow turns round inside the Hill sphere and goes back out (the horseshoe region).
\citet{lubow99} showed that the flow distributed in a certain band (see, Figure~4 of \citealt{lubow99}) accretes onto the protoplanetary system in a two-dimensional simulation \citep[see also,][]{tanigawa02a}.
However, in three-dimensional simulations, only a part of the flow in the band (or bundle) in the radial direction flows into the protoplanetary system, as shown in Figures~\ref{fig:6} and \ref{fig:7}.
\citet{kley01} and \citet{paardekooper08} pointed out that the mass accretion rate derived in three-dimensional calculations is much smaller than that derived from two-dimensional calculations.

As shown in Figure~\ref{fig:6}, the bandwidth that limits the gas flowing into the protoplanetary system is proportional to {\it min}\,($r_B$, $\rh$).
The Bondi radius is proportional to $\propto M_{\rm p}^2$ (eq.~\ref{eq:bondi}), while the Hill radius is proportional to $\propto M_{\rm p}^{1/3}$ (eq.~\ref{eq:hill}).
When the Bondi radius is smaller than the Hill radius ($r_{\rm B}<\rh$), the bandwidth ($\sim r_{\rm B}$) and accretion rate rapidly increase with the protoplanet mass because the mass dependence of the Bondi radius is strong ($r_{\rm B}\propto M_{\rm p}^2$).
On the other hand, when $\rh > r_{\rm B}$, the bandwidth ($\sim \rh$) and accretion rate maintain  an almost constant value, because the mass dependence of the Hill radius is considerably weaker ($\rh\propto M_{\rm p}^{1/3}$).
It should be noted that the Bondi radius equals the Hill radius when $M_{\rm p} = 0.023\, (\ro/1{\rm AU})^{3/4}\mj$.
Thus, as shown in Figure~\ref{fig:5}, it is possible to qualitatively understand the protoplanet mass dependence on the accretion rate: the accretion rate rapidly increases when $r_{\rm B}< \rh$, while it maintains an almost constant value when $r_{\rm B} > \rh$.
However, the power of the accretion rate ($\dot{M}_{\rm p}\propto M^{1.5}$ for $r_{\rm B}<\rh$, see Fig.~\ref{fig:5}) derived in our calculations is slightly smaller than the Bondi solution \citep{bondi52}: 
\begin{equation}
\dfrac{dM_{\rm p}}{dt} =  \dfrac{4\,\pi\, G^2\, M_{\rm p}^2 \rhoz}{c_{s,0}^3} \propto M_{\rm p}^2.
\end{equation}
This discrepancy can be attributed to the fact that spherically symmetric flow in an isolated system is assumed in Bondi accretion.
However, in the protoplanetary system, the gravitational sphere is limited to the Hill sphere, and the angular momentum that was acquired from the shearing motion in the protoplanetary disk can affect the gas accretion. 
Thus, it is natural that the power of the accretion rate in the protoplanetary disk does not completely match with that obtained from the Bondi solution.
In fact, the accretion rate derived in this simulation is two or three orders of magnitude smaller than the Bondi accretion rate.
As has been mentioned, the Bondi radius is useful in representing the gas flow when the protoplanet is less massive, indicating that the thermal pressure cannot be ignored when the thermal energy dominates the gravitational energy inside the Hill sphere. 
Note that, in the isothermal model, the thermal energy dominates the gravitational energy inside the Hill sphere when $M_{\rm p} < 0.08\mj$  \citep[for details see][]{machida08}.

\subsection{Gas Flow around a Protoplanet}
\label{sec:small}
In this subsection, we consider the gas flow patterns near the protoplanet.
Figure~\ref{fig:8} shows the density distributions and streamlines in the neighborhood around the protoplanet for models M005A2 (left) and M1A2 (right).
The streamlines are inversely integrated from the high-density region (yellow constant density surface).
The left panel in Figure~\ref{fig:8} shows that the gas spirals into the protoplanet in the vertical direction.
On the other hand, in the right panel, a part of the gas accretes onto the protoplanet, while the remainder accretes onto the circumplanetary disk.
The gas accreting onto the circumplanetary disk orbits in the disk and gradually falls into the protoplanet. 
The density contours in wall surfaces show a thick torus-like configuration of the disk in model M005A2, and a very thin disk in model M1A2.

Figure~\ref{fig:9} shows the mass flux ($\rho \vert v_r \vert$) for model M005A2, in which only mesh points having $v_r<0$ are plotted in color.
Thus, the gas in the black region has a positive radial velocity $v_r>0$ and flows out from the protoplanetary system.
Figures~\ref{fig:9}{\it a} and {\it b} indicate that a large part of the gas enters into the protoplanetary system from the vertical direction around the Hill sphere.
However, the gas does not enter into the protoplanetary system on the equatorial plane.
Figure~\ref{fig:9}{\it c} shows that, even when the gas has a negative radial velocity $v_r<0$ inside the Hill sphere ($\tl{r}<\rht$), it cannot enter into the region $\tl{r}\lesssim \rht/2$ on the equatorial plane: the flow turns round and returns (see, for example, Figure~\ref{fig:7}).
Therefore, although the gas enters the Hill sphere, it cannot reach the protoplanet on the equatorial plane.
\citet{paardekooper08} also showed that the gas flows out from the Hill sphere on the equatorial plane (the equatorial outflow).

\citet{machida08} showed that the specific angular momentum of the protoplanetary system increases in proportion to $j\propto M_{\rm p}$.
They also showed that the circumplanetary disk appears when the gravitational energy exceeds the thermal energy in the whole region of the Hill sphere, which occurs when $M_{\rm p}>0.08\mj$ at the Jovian orbit.
Thus, when the protoplanet mass exceeds $M_{\rm p}>0.08\mj$ at the Jovian orbit, the circumplanetary disk appears, and then it increases its size and mass with the protoplanet mass because the massive protoplanet acquires a larger amount of angular momentum from the shearing motion in the protoplanetary disk.
One can expect that the emergence of the circumplanetary disk may reduce the accretion rate.
However, we confirmed that the circumplanetary disk keeps a constant mass, indicating that the circumplanetary disk does not strongly affect the gas accretion rate onto the protoplanet.
In summary, although the flow pattern around a massive protoplanet is different from that around a less massive protoplanet, fine structures such as the circumstellar disk around the protoplanet does not strongly affect the accretion rate onto the protoplanet.
Thus, once the gas is gravitationally captured by the protoplanet, it accretes onto the protoplanet.

\subsection{Thermal Effects on the Mass Accretion Rate}
\label{sec:thermal}
As shown in Figure~\ref{fig:5}, when the protoplanet is less massive ($M\lesssim 0.2\mj$), the accretion rates for models with different equations of state have almost the same value.
On the other hand, when the protoplanet mass exceeds $M_{\rm p}\gtrsim 0.2\mj$, accretion rates are slightly different for models with different equations of state.
This slight difference is considered to be caused by thermal evolution around the protoplanet.
The deviation from an isothermal approximation is small in a less massive system, while it becomes slight large in a massive system, as shown in \citet{machida09}.
The density distribution around the protoplanet for models M1I (left) and M1A1 (right) is shown in Figure~\ref{fig:10}, in which the circumplanetary disks are represented by constant density surfaces.
The figure indicates that the isothermal model (model M1I) has a thinner disk than the adiabatic model (model M1A1).
The orange constant density surface of $\tl{\rho}=10^2$ shows that model M1A1 has a torus-like disk that flared up outwardly, while model M1A1 has a thick disk.
The red constant density surface of $\tl{\rho}=10^3$ shows that, near the protoplanet,  model M1A1 has a more compact disk than model M1I.

Figure~\ref{fig:11} illustrates the mass of the gas envelope (or the circumplanetary disk), which is integrated from the centre.
The figure indicates that a model with a harder equation of state has a more massive envelope.
Since the circumstellar disk in these models is sufficiently gravitationally stable (Toomre's Q $\gg 1$), the gas in the circumplanetary disk barely accretes onto the protoplanet with dynamic instability.
However, as shown in Figure~\ref{fig:10}, the model with large thermal energy, that is, a harder equation of state, has a more massively flared disk.
In such a disk, the effect of centrifugal and gravity forces is relatively small, because the thermal energy is relatively large.  
Therefore, it is likely that the accretion rate in such a disk becomes higher than that in a model with lower thermal energy in which the circumstellar disk is strongly supported by the centrifugal force, because the path of the gas streamlines accreting onto the protoplanet is changed.
Finally, it follows that the massive disk has a larger accretion rate.

At last, we comment on a mass fraction of the adiabatic gas.
For less massive protoplanets, owing to the shallow gravitational potential, the density enhancement is small even near the protoplanet.
Since the gas density is connected to the gas temperature through the equation of state (eq.~[\ref{eq:eos}]), such models may have a small fraction of the adiabatic gas.
We estimated a mass fraction of the adiabatic gas for model with $\rht=0.29$ ($M_{\rm p}=0.01\mj$) and $\rhoc=\rho_0$ (model M001A0).
The mass fraction $f$ of the adiabatic gas to the total mass within the Hill 
($r < \rh$), half Hill ($r<1/2\, \rh$) and 1/10 Hill ($r<1/10\, \rh$) radii are $f$ = 0.26, 0.55 and 1.0, respectively.
In addition, the fraction inside the Bondi radius ($r < r_{\rm B} =0.25\rh$) is $f$ = 0.69.
As shown in \S\ref{sec:large}, the Bondi radius is more important to investigate the gas accretion when $r_{\rm B}<\rh$. 
A large fraction of total mass is adiabatic gas inside the Bondi (or Hill) radius, while there is small difference in the mass accretion rate among isothermal and adiabatic models, as seen in Figure~\ref{fig:5}.
As a result, we concluded that the mass accretion rate barely depends on the thermodynamics around the protoplanet.

\section{Comparison with Previous Simulations}
\label{sec:discuss}
The mass accretion rate onto the protoplanet (system) in the protoplanetary disk was investigated in some previous studies \citep{kley01,tanigawa02a,dangelo03,bate03,klahr06,dobbs07,paardekooper08,fouchet08,dangelo08,ayliffe09}.
Subsections \S\ref{sec:two-iso} to \ref{sec:three-rad} will explain the classification of the previous simulations by classifying them into three categories and discussing the salient features of each category.

\subsection{Two-dimensional Isothermal Calculations}
\label{sec:two-iso}
\citet{tanigawa02a} derived the mass accretion rate under the isothermal gas condition in a two-dimensional local simulation.
The spatial resolution of their calculation is comparable to our study.
However, the mass accretion rate is greatly different from the results obtained in this study.
In \citet{tanigawa02a}, the mass accretion rate continues to increase as a function of the protoplanet mass, while it saturates at $M_{\rm p}\simeq0.2\mj$ in our three-dimensional calculation.
Furthermore, although in this study, the mass accretion rate increases with the protoplanet mass in the range of $M_{\rm p}\lesssim 0.2\mj$, its value is much smaller than in \citet{tanigawa02a}.
For example, at $M_{\rm p}=0.1\mj$, the accretion rate in \citet{tanigawa02a} is $\dot{M} = 2.3\times10^{-3}\mj$\,yr$^{-1}$ (see, Equation~[19] of \citealt{tanigawa02a}), while it is  $\dot{M}=2.7\times 10^{-5}\mj$\,yr$^{-1}$ in our isothermal model.
Thus, in the range of $M_{\rm p}\lesssim 0.2\mj$, the accretion rate based on a two-dimensional calculation is two orders of magnitude larger than that obtained from a three-dimensional calculation.
The difference becomes larger in the range of $M_{\rm p}>0.2\mj$, because the accretion rate continues to increase in two-dimensional calculations, while it gradually decreases in the three-dimensional calculations.
\citet{kley01} compared the mass accretion rate in their three-dimensional simulations with that in two-dimensional simulations \citep{kley99} and showed that the mass accretion rate derived in three-dimensional simulations is smaller than that in two-dimensional simulations.
\citet{paardekooper08} also commented on the decrease in mass accretion observed in three-dimensional simulations compared with two-dimensional simulations. 
The difference in mass accretion is considered to be caused by the spatial dimensions.

\subsection{Three-dimensional Isothermal Calculations}
\label{sec:three-iso}
\citet{kley01}, \citet{dangelo03}, and \citet{bate03} estimated the mass accretion rate under the locally isothermal assumption in global simulations.
The mass accretion rates derived from these global simulations corresponds well to the rate derived from our local simulations, despite the fact that the spatial resolution and size of the sink are considerably different.
For example, \citet{dangelo03} estimated the mass accretion rate in their global three-dimensional simulations with a spatial resolution of $\Delta{\tl{x}} = 0.06\,\rht$, which is 40 times coarser than ours, and defined the sink as the region inside one-tenth of the Hill radius ($\rs<0.1\,\rht$), which is 28 times larger than our sink size (see \S\ref{sec:sink}). 
Thus, our study can resolve the present Jovian radius, while \citet{dangelo03} cannot resolve the radius.
As well, in their calculation, the circumplanetary disk cannot be sufficiently resolved because the circumplanetary disk is formed in the region $r<10-50\,\rp$ \citep{machida09}.
In summary, the differences between \citet{dangelo03} and our study are the spatial resolution and numerical setting. 
We calculated the evolution of the protoplanetary system in the local simulation with finer spatial resolution, while this cannot be done in \citet{dangelo03}'s global simulation with coarser spatial resolution.
Despite the differences, the accretion rate in our isothermal models corresponds well to their results (compare Figure \ref{fig:5} with Figure~7 of \citealt{dangelo03}).
The mass accretion rate has a peak value of $\dot{M}\sim 8\times10^{-5}\mj$\,yr$^{-1}$ at $M_{\rm p} = 0.2-0.3\mj$ in \citet{dangelo03}, while it has a peak value of $\dot{M}\sim4\times10^{-5}\mj$\,yr$^{-1}$ at $M_{\rm p} \simeq 0.2\mj$ in our study.
Furthermore, the mass accretion rates at $M=0.01$ and $1\mj$ are $\dot{M} \sim 3\times10^{-7} \mj$\,yr$^{-1}$ and $\sim4\times 10^{-5}\mj$\,yr$^{-1}$ in \citet{dangelo03}, while they are $8.2\times 10^{-7}\mj$\,yr$^{-1}$ and  $3.6\times10^{-5}\mj$\,yr$^{-1}$ in our study.
It should be noted that although \citet{dangelo03} investigated the evolution of the protoplanetary disk by changing the gravitational potential of the protoplanet, the mass accretion rates differed slightly in each model.

\citet{kley01} investigated the evolution of the protoplanetary disk only when the protoplanet has a Jovian mass ($1\mj$) under the isothermal gas condition in their three-dimensional simulations, in which the spatial resolution is coarser than \citet{dangelo03}.
They found that a mass accretion rate of $6\times 10^{-5}\mj$\,yr$^{-1}$,  while it is $(2.2-4.1)\times10^{-5}$\,yr$^{-1}$ at $1\mj$ in our  study (see, for example, Fig.~\ref{fig:5}).
Thus, our results correspond well with their results.
\citet{bate03} also calculated the evolution of the protoplanetary disk in a three-dimensional global simulation with a spatial resolution comparable to that of \citet{dangelo03}.
The accretion rate derived in \cite{bate03} corresponds well to ours within a factor of three.
In addition, it shows the same trends as in our study.
That is, the accretion rate has a peak at $M_{\rm p}\simeq 0.1\mj$ and decreases for $M_{\rm p}>0.1\mj$.

In summary, all three-dimensional calculations show almost the same mass accretion rate, though they have adopt different calculation settings (i.e., local or global calculation), spatial resolutions and sizes of the sink radius.
In these calculations, the Hill radius or the region near the Hill sphere is resolved, while the region near the protoplanet, much smaller scale than the Hill radius is not always resolved sufficiently. 
This indicates that the mass accretion rate is regulated by the region around the Hill sphere, not by the details at smaller scale length.
Thus, we conclude that the mass accretion rate is safely estimated when the Hill radius is resolved with adequate spatial resolution.

\subsection{Three-dimensional Radiative Calculations}
\label{sec:three-rad}
Recently, \citet{klahr06}, \citet{paardekooper08}, \citet{fouchet08} and \citet{ayliffe09} investigated the evolution of the protoplanetary disk using three-dimensional, radiation-hydrodynamical simulations.
\citet{klahr06} estimated the mass accretion rate of $5.1 \times 10^{-5}\mj$\,yr$^{-1}$ when the protoplanet has a Jovian mass.
They concluded that the mass accretion rate in a radiative model is larger than that in an isothermal model.
\citet{paardekooper08} calculated the evolution of the protoplanetary disk with several models with different protoplanetary mass under the locally isothermal approximation, while they calculated it with the radiation-hydrodynamical simulations only with protoplanet mass of $0.6\me$ and $5\me$.
The accretion rates in their isothermal calculation are identical to our results.
As well, the accretion rate in their radiation-hydrodynamical simulations is also comparable to our results.
It should be noted that, since the range of the protoplanetary mass in their radiation-hydrodynamical simulations is different from our study, a direct comparison of the results cannot be performed.
They concluded that although the accretion rate in the radiation-hydrodynamical simulation is smaller than that in the isothermal simulation, the difference is not dramatic.
It should be noted that, for radiation-hydrodynamical simulations, \citet{paardekooper08} only calculated the disk evolution with very small protoplanetary masses of $0.6\me$ and $5\me$. These masses are considerably different from that used by \citet{klahr06}.

\citet{fouchet08} investigated the evolution of the protoplanet disk including both the self-gravity and radiation physics for a protoplanet with a Jovian mass, $1\mj$.
They also calculated the protoplanetary disk under the isothermal approximation and compared them with the radiation-hydrodynamical simulation.
They obtained an accretion rate of $5\times10^{-5}\mj$\,yr$^{-1}$ in the isothermal simulation and $\sim2\times10^{-5}\mj$\,yr$^{-1}$ in the radiation-hydrodynamical simulation.
These rates are well identical to  \citet{klahr06} and our results of $(2.2-4.1)\times10^{-5}$\,yr$^{-1}$ at $M_{\rm p}=1\mj$.
\citet{ayliffe09} also estimated the mass accretion rate in their radiation-hydrodynamical simulation and found an accretion rate similar to those derived in previous studies.

However, the radiative effect on the mass accretion rate is controversial.
\citet{paardekooper08} and \citet{fouchet08} showed that the accretion rate in a nonisothermal disk is reduced compared to the isothermal disk \citep[see also,][]{ayliffe09}, while \citet{klahr06} showed the mass accretion rate in radiative model is larger than in the isothermal model.
In our study, the accretion rate for the isothermal model is smaller than for the adiabatic models.
However, the difference in the accretion rate between the isothermal and radiative models is not so large.
This can be attributed to the gravitational energy of the protoplanet dominating the thermal energy \citep{machida09}.
Thus, it is expected that the accretion rate barely depends on the thermal evolution in the protoplanetary disk.
For example, when the Jovian mass planet is adopted at Jovian orbit, the accretion rate is in the range $(2-6)\times10^{-5}\mj$\,yr$^{-1}$ \citep{kley01,dangelo03,bate03,klahr06,fouchet08,dangelo08}. 
This indicates that the growth timescale of the Jovian planet at Jovian orbit is $\sim10^5$\, years.

\subsection{Disk Viscosity and Effect of the Gap on the Mass Accretion Rate}
\label{sec:gap}
In this subsection, we discuss the relation between gap formation and the mass accretion rate onto the protoplanet.
Some previous studies pointed out that the reduction of the mass accretion rate for massive protoplanets is related to the gap formation \citep[e.g.,][]{dangelo03,bate03}.
The density gap in the protoplanetary disk is the result of the competition between torques exerted on the disk by the planet and by the disk itself.
The planet gives angular momentum to the outer part of the disk, and it takes angular momentum from the inner part of the disk \citep{goldreich80}.
Thus, a protoplanet tends to open a gap.
On the other hand, the disk kinematic viscosity makes the gas back to the gap.
As a result, the gap formation depends on both the disk kinematic viscosity and mass of the protoplanet \citep{lin86,bryden99}.
A massive protoplanet in the disk with a smaller viscosity tends to show a clear gap.
However, the formation condition, size, and width of the gap are not clearly understood.

In the inviscid ($\nu=0$) gas disk model as adopted in this study, it is considered that the gap formation occurs when the Hill radius exceeds the scale height of the protoplanetary disk, i.e., $\rh \gtrsim h$ \citep{lin86,bryden99,crida06}.
Using equations~(\ref{eq:negular-scale-height}) and (\ref{eq:hill}), this condition can be described as
\begin{equation}
\dfrac{M_{\rm p}}{M_{\rm Jup}} \gtrsim 0.12 \left( \dfrac{\ro}{1\,{\rm AU}}\right)^{3/4},
\label{eq:gap2}
\end{equation}
which  indicates that, at Jovian orbit, the density gap begins to appear when the mass of the protoplanet exceeds $M_{\rm p} \gtrsim 0.4 \mj$.
As shown in \S\ref{sec:large}, we found that the gas flows into the planetary system in the region $x=1-3\,\rh$ when the protoplanet is more massive than $M_{\rm p}> 0.08\,\mj$ at Jovian orbit.
Thus, when the gap width $\Delta_{\rm gap}$ becomes larger than $\Delta_{\rm gap} \gtrsim 2-6\,\rh$, the mass accretion rate onto the protoplanet is expected to decline.
The decline of the mass accretion rate at $M_{\rm p}\gtrsim 0.3-1\mj$ in global simulations \citep[e.g.,][]{dangelo03} is caused by a wide gap formation.
On the other hand, our local simulation shows no clear gap, though the low-density region is shallower than that in the global simulations.
Local simulations are not appropriate to treat gap formation because of the radial boundary condition \citep{miyoshi99,tanigawa02a}, since the gap properties, such as the gap depth and width, depend on the size of the simulation box. 
Therefore, when the mass of protoplanets exceeds $M_{\rm p} \gg 0.4\mj$ at Jovian orbit, it is expected that the mass accretion rate derived in the local simulation is smaller than that in the global simulation owing to the gap.
On the other hand, in the range of $M_{\rm p} \lesssim 0.4\mj$, the mass accretion rate derived in the local simulation is always applicable, because no clear gap forms in this mass range.
In addition, the mass accretion rate derived in this study would be valid even for $M_{\rm p}>0.4\mj$ when the disk viscosity is sufficiently large, because no clear gap appears in such a viscous disk.
Moreover, the mass accretion rate in the local simulation well agrees with that in the global simulation in the range of $M_{\rm p}\le 1\mj$ as denoted in \S\ref{sec:three-iso}.
This may be because the density gap barely affects the gas accretion in this mass range.
However, global simulations show a rapid decline of the mass accretion rate for $M_{\rm p}>1\mj$, in which the density gap seems to significantly affect the mass accretion. 
In such a case, the mass accretion rate derived in the local simulation cannot be applicable.

In this study, the mass accretion rate is derived in the local simulation.
This accretion rate may be overestimated when the protoplanet is sufficiently massive and the protoplanetary disk has a sufficiently small viscosity, because the gap formation decreases the mass accretion rate.
Thus, the mass accretion rate obtained here corresponds to the possible maximum value that is attainable without forming a density gap in the disk. 
However, even when the density gap is formed, the formula of the mass accretion rate is applicable with the reduced disk surface density owing to the gap, because we investigated the gas flow in a dimensionless form, and all physical quantities are scalable, as denoted in Equation~(\ref{eq:mdot2}).

\section{Growth Timescale of Gas Giant Planets}
As mentioned in \S\ref{sec:intro}, to form a gas giant planet, the solid core with a mass of $1-10\,M_\oplus$ needs to acquire the gas component from the protoplanetary disk.
At the end of the ``oligarchic growth'' stage \citep{kokubo98}, several protoplanets composed of the solid component, which will be referred to as the planetary core, appear with an orbital separation of about 10 Hill radii \citep{kokubo02}.
Only when a planetary core acquires sufficient gas from the protoplanetary disk before the dissipation of the protoplanetary gas disk, the gas giant planet is possible to
form.
The mass and growth timescale of the planetary core were investigated in detail using $N$-body simulations.
In the remainder of this section, using the mass accretion rate onto the gas giant planet obtained in this study, as well as the mass and growth timescale of the planet core obtained from $N$-body simulations \citep{kokubo02}, an investigation of the final mass and growth timescale of the gas giant planet is performed.

\citet{kokubo02} derived an (isolation) mass and growth timescale of the planetary core  as a function of disk model.
Adopting $\tl{b}=10$ (the orbital separation normalized by the mutual Hill radius of the cores), $\alpha=3/2$ (the power index of the radial surface density distribution), and $M_* = \msun$ (the mass of the central star) in Equations~(17) and (26) of \citet{kokubo02},  the isolation mass $M_{\rm iso}$ and growth timescale $\tau_{\rm core}$ of the planetary core can be described as
\begin{equation}
M_{\rm iso} \simeq 0.16  \left( \dfrac{f_{\rm ice}\, \Sigma_1}{10} \right)^{3/2} \left( \dfrac{a_{\rm p}}{1 {\rm AU}} \right)^{2} 
 M_\oplus,
\label{eq:miso}
\end{equation}
\begin{equation}
\tau_{\rm core} \simeq 3.2 \times 10^5 \, f_{\rm ice}^{-1/2} \left( \dfrac{\Sigma_1}{10}  \right)^{-9/10}
\left( \dfrac{a_{\rm p}}{1 {\rm AU}} \right)^{59/20} \ \ {\rm yr},
\label{eq:tgrow}
\end{equation}
where $f_{\rm ice}$ (ice factor) and $\Sigma_1$ (reference solid surface density at 1\,AU) can be related to the solid surface mass density as
\begin{equation}
\Sigma_{\rm solid} = f_{\rm ice}\, \Sigma_1 \left( \dfrac{a_{\rm p}}{1 {\rm AU}} \right)^{-3/2} \ \ \ {\rm g\ cm}^{-3}.
\end{equation}
The ice factor $f_{\rm ice}$  expresses the increase of solid by ice condensation over the snow-line $a_{\rm snow}$ and $f_{\rm ice}=1$ ($a<a_{\rm snow}$=2.7\,AU) and 4.2 ($a>a_{\rm snow}$).
As the reference solid surface density, we adopt the minimum-mass disk model \citep{hayashi85} with $\Sigma_1 = \Sigma_{\rm 1}^{\rm H} = 7$.
In the following, the solid surface density is parameterized as $\Sigma_1/\Sigma_{\rm 1}^{\rm H}$ = 1/2, 1 and 2.
We set the mass of the planetary core as the isolation mass $M_{\rm core}=M_{\rm iso}$.

In a conventional scenario of a gas planet formation \citep[e.g.,][]{mizuno80, stevenson82, bodenheimer86}, the planetary core is heated by the continuous accretion of planetesimals, and has a hydrostatic gas envelope when a planetary core is less massive than $M_{\rm core} \lesssim 10\me$.
Then, the planetary core cannot sustain the gas envelope and the rapid gas accretion is triggered (i.e., the runaway gas accretion phase) after it grows up to $M_{\rm core} \gtrsim 10\me$. 
However, the current planetary accretion theory \citep[e.g.,][]{kokubo98,kokubo02} indicates that the accretion of planetesimals almost halts after the isolation core formation (or after the oligarchic growth stage).
In this case, the planetary core cannot have a hydrostatic envelope owing to the absence of the heat source even when the planetary core is $M_{\rm core}\lesssim 10\me$.
Therefore,  the runaway gas accretion is triggered just after the isolation core formation.

Figure~\ref{fig:12}{\it a} shows the mass of the planetary core  for different $\Sigma_1$ based on Equation~(\ref{eq:miso}).
In the figure, the sudden rise of the core mass at 2.7\,AU is caused by the increase of solid component at the snow-line.
In the standard model ($\Sigma_1 = \Sigma_{\rm 1}^{\rm H}$, the red line), the planetary core  has a mass of $2.8\,\me$ at the Jovian orbit ($5.2$\,AU) and $4.4\,\me$ at the Saturnian orbit ($9.6$\,AU).

Figure~\ref{fig:12}{\it b} shows the growth timescales of the planetary core $\tau_{\rm core}$ (thin solid lines, Equation~[\ref{eq:tgrow}]) and Jovian-mass gas planet $\tau_{\rm gas}$ (thick solid lines) in each orbit.
Using Equation~(\ref{eq:mdot}), the mass accretion rate is integrated as a function of time until the planetary core reaches the Jovian mass $\mj$ in each orbit using
\begin{equation}
M_{\rm Jup} = M_{\rm core} + \int_{0}^{\tau_{\rm gas}}  \dfrac{dM_{\rm p}}{dt} dt, \ \ M_{\rm p}(t =0)= M_{\rm core},
\label{eq:mass-int}
\end{equation}
where $L=L_\odot$, $T=T_0$ and $\rho_0 = \rho_0 (\Sigma_1/\Sigma_{\rm 1}^{\rm H})$ are used in Equation~(\ref{eq:mdot}).
Thus, $\tau_{\rm gas}$ represents the time required to form a Jovian-mass gas  planet after the formation of the isolation core.
The figure indicates that $\tau_{\rm gas} \ll \tau_{\rm core}$ in any orbit.
For example, in the standard model, that is, $\Sigma_1 = \Sigma_{\rm 1}^{\rm H}$, the growth timescale of the planetary core is $\tau_{\rm core}=2.7\times 10^7$\,yr at the Jovian orbit and $1.7\times10^8$\,yr at the Saturnian orbit, while that of a Jovian gas planet is $\tau_{\rm gas} = 4.7\times 10^4$\,yr at the Jovian orbit and $7.7\times10^{4}$\,yr at the Saturnian orbit.
Therefore, the gas-accretion timescale is two or three orders of magnitude shorter than the core growth timescale.

As shown in \citet{kokubo02}, gas giant planets can form in the limited range of the protoplanetary disk where the planetary core is large enough to capture the gas component before the dissipation of the gas disk.
The solid curves of Figures~\ref{fig:12}{\it c} and {\it d} show the planet mass for the disk lifetime of $t_{\rm gas}=5\times10^4$\, yr ({\it c}) and $10^6$\,yr ({\it d}).
In these figures, using Equation~(\ref{eq:mdot}), the planet mass $M_{\rm p}$ at each orbit is derived by
\begin{equation}
M_{\rm p} = M_{\rm core} + \int^{t_{\rm gas}}  \dfrac{dM_{\rm p}}{dt} dt, \ \ M_{\rm p}(t =0)= M_{\rm core}.
\label{eq:final-mass}
\end{equation}
The dotted lines of Figures~\ref{fig:12}{\it c} and {\it d} indicate the maximum mass in the ring of the protoplanetary disk around the planet \citep{hayashi85}, which is described as
\begin{equation}
M_{\rm max} = 2 \pi \ro\, \delta r \,\Sigma_{\rm g},
\label{eq:mmax}
\end{equation}
where $\delta r$ is the ring width and is expressed by $ \delta r = 2 f_{\rm H}\, \rh$, where $f_{\rm H}$ is the parameter representing the ring width.
Adopting $M_{\rm p} = M_{\rm max}$ and $M_{\rm c} = \msun$ in equation~(\ref{eq:hill}), $\delta r$ can be described as 
$ \delta r = 2 f_{\rm H}\, \ro ( M_{\rm max}/{3 \msun}  )^{1/3}. $
Thus, Equations~(\ref{eq:mmax}) gives
\begin{equation}
M_{\rm max} = \left( 4 \pi\, \ro^2\, f_{\rm H} \Sigma_{\rm g} \right)^{3/2} \left({3 \msun} \right)^{-1/2} .
\label{eq:mmax2}
\end{equation} 
In equation~(\ref{eq:mmax2}), $\Sigma_{\rm g} = f_g\, \Sigma_{\rm g}^{\rm H}$ is adopted as the gas surface density of protoplanetary disk, where $f_{\rm g}$ is related to the solid surface density as $f_{\rm g} = \Sigma_1/\Sigma_{\rm 1}^{\rm H}$ because the constant ratio of the gas to solid surface density is assumed.
The fiducial gas surface density $\Sigma_{\rm g}^{\rm H}$ is given by the standard disk model \citep{hayashi85} as
\begin{equation}
\Sigma_{\rm g}^{\rm H} = 1.7\times10^3 \left( \dfrac{\ro}{1\,{\rm AU}} \right)^{-3/2}\ \  {\rm g}\,{\rm cm^{-2}}.
\end{equation}
Equation~(\ref{eq:mmax2}) is the mass of the ring in the protoplanetary disk in the range of $\vert x \vert < f_{\rm H}\,\rh$, where the origin of $x$ is the position of the planet.
As shown in \S\ref{sec:large}, our calculation indicates that the gas in the range of $\vert x \vert \lesssim 3\,\rh$ flows into the protoplanet system.
Thus, $f_{\rm H}=3$ is adopted in equation~(\ref{eq:mmax2}).
Since the protoplanet cannot acquire the gas component exceeding the mass in this ring, the mass $M_{\rm max}$ means the maximum mass of the gas component acquired by the protoplanet at each orbit \citep[for details see,][]{hayashi85}.

The thick gray curves of Figures~\ref{fig:12}{\it c} and {\it d} are the planet mass attainable at each orbit.
Figure~\ref{fig:12}{\it c} indicates that the planet mass is determined by the dotted line (i.e., eq.~[\ref{eq:mmax2}]) for $\ro < 5$\,AU, while by the solid curve (i.e., eq.~[\ref{eq:final-mass}]) for $\ro > 5$\,AU.
For $\ro<5\,$AU, since the growth timescale of the gas planet is shorter than the lifetime of the gas disk, the planetary core accumulates all gas in the ring with  $\vert x \vert <3\rh$ in $t<5\times10^4$\,yr.
On the other hand, for $\ro>5$\,AU, since the gas disk dissipates before the protoplanet accumulates all gas in the ring, the planet mass is determined by equation~(\ref{eq:final-mass}).
As shown in equation~(\ref{eq:mdot}), since the mass accretion rate is a decreasing function of the orbital radius $\ro$, the planet mass decreases with $\ro$ for $\ro>5$\,AU.
The thick gray curve in Figure~\ref{fig:12}{\it c} shows that the planet, respectively, has almost the Jovian and Saturnian mass at Jovian and Saturnian orbit.
For $\ro > 10$\,AU, the mass of the gas component is less massive  and comparable to the solid core mass ($M_{\rm p}\sim M_{\rm core}\sim 10\me$).
In addition, the gas planet formed at $\ro \gtrsim 2.7$\,AU has a core mass of $1\me<M_{\rm core}<10\me$.
Note that \citet{saumon04} studied the internal structure of Jupiter and Saturn, and estimated the core mass of the Jupiter in the range of $0-11\me$, and core mass of the Saturn in the range of $9-22\me$.
These features (mass of gas and solid components at each orbit) well correspond to those of giant planets in our solar system (Jupiter, Saturn, Uranus, and Neptune).

Although we assumed that the gas disk dissipates in $t=5\times10^4$\,yr after the isolation core formation in Figure~\ref{fig:12}{\it c}, observations indicate that the protoplanetary disk has a lifetime of $\simeq 0.3-30$ Myr \citep{haisch01,hartmann05,silverstone06,flaherty08}.
Figure~\ref{fig:12}{\it d} shows the planet mass when the gas disk exists for $t=10^6$\,yr after the isolation core formation.
In this figure, the planet mass is determined only by the dotted line (i.e., the maximum mass in the ring, eq.~[\ref{eq:mmax2}]).
As shown in Figure~\ref{fig:12}{\it b}, since the growth timescale of a gas planet is shorter than $10^6$\,yr at any orbit, the protoplanet can accumulate all gas component around it when the gas disk exists for $t>10^6$\,yr.
The thick gray line in Figure~\ref{fig:12}{\it d} shows that although the Jovian mass planet is formed at Jovian orbit, more massive planets than Jovian mass are formed at Saturnian, Uranian, and Neptunian orbits, which contradicts our solar system.

Figures~\ref{fig:12}{\it c} and {\it d} indicate that the gas giant planet with the core mass of $M_{\rm core}>1\me$ can be formed in the orbital range of $\ro>2.7$\,AU.
They show that the mass of planets exceeds Jovian mass in the range of $\ro \gtrsim 5$\,AU when the gas disk exists for $t_{\rm gas}>10^5$\,yr after the isolation core formation.
Thus, to realize our solar system in the present model, it is expected that the gas component of the protoplanetary disk needs to dissipate just after the isolation core formation.

We need to pay attention to viscous evolution of the disk for estimating the planet mass.
In Figures~\ref{fig:12}{\it c} and {\it d}, the planet mass in the range of $\ro<2.7$\,AU for disk with $\tau_{\rm gas}=5\times10^4$\,yr and in whole orbital range for disk with $\tau_{\rm gas}=10^6$\,yr is determined by the disk surface density at the location of the planetary core formation when the gas accretion timescale is shorter than the disk lifetime.
In other words,  the planet mass is determined by in situ surface density (or in situ disk mass).
However, in these figures, the disk viscous evolution is ignored.
If the gap is refilled by the viscous evolution, the planet can acquire the gas component exceeding $M_{\rm max}$.
Thus, the planet mass becomes more massive than gray line in these figure, when the disk has a significantly large viscosity.

At last, we comment on the hydrostatic envelope.
As shown in the above, to estimate the formation time of a gas giant planet, we assumed no planetesimal accretion onto the planetary core after the isolation core formation.
Although this assumption is supported by recent numerical simulations, the formation timescale of a gas giant planet may be considerably different when the planetesimal accretion does not completely halt even after the isolation core formation.
With the planetesimal accretion, the planetary core can sustain a hydrostatic envelope for $\sim10^8$\,yr at the maximum \citep{ikoma00}.
In this situation, the formation time (or gas accretion time) for the gas giant planet is prolonged.
Thus,  $\tau_{\rm gas}$ in Figure~\ref{fig:12}{\it b} gives the minimum time necessary for the formation of gas giant planets.

\section{Summary and Discussion}
In this study, a calculation of gas accretion onto a planetary core including the thermal effect was performed, in which the mass accretion rate and growth timescale of a gas giant planet were estimated.
l,

The mass accretion rate derived in this study quantitatively agrees with those derived in previous simulations.
It is surprising that the same value for the accretion rate is derived from three-dimensional simulations with different parameter values, including parameters such as spatial resolution, the size of the sink, and the treatment of the thermal effect, including locally isothermal approximation, barotropic equation of state, and radiative hydrodynamics.
This suggests that the mass accretion rate is only determined by `the protoplanet mass'  and `the properties of the protoplanetary disk.'
In our calculation, we resolved the Jovian radius of the cell width.
On the other hand, for example, \citet{kley01} derived the mass accretion rate by using the sink radius of $\rh/2$.
In \S\ref{sec:small}, we showed that almost all gas falling into the half Hill radius ($r < \rh/2$) can accrete onto the surface of the protoplanet.
Although the radiative effect can affect the mass accretion rate, the difference is less than a factor of $\sim3$.
As a result, this study, like previous studies, shows that the mass accretion rate can be accurately estimated for resolution higher than half Hill radius.
In other words, the high spatial resolution that resolve Jovian radius in not necessary to investigate the mass accretion rate and growth times scale of gas giant plants.

A comparison of our results with the $N$-body simulations for the solid core aggregation shows that the gas giant planet can form in the orbital range of $\ro > 2.7\,{\rm AU}$ in a short timescale of $\sim10^4-10^5$\,yr after the planetary core inside the gas giant planet is formed in a longer timescale of $\sim10^8$\,yr.
In each orbit, the growth timescale of the gas giant planet is two orders of magnitude shorter than the core aggregation timescale.
This may indicate that, in our solar system, the planetary core with $1-10\me$ began to capture the gas component to form gas giant planets such as Jupiter and Saturn just before the dissipation of the protoplanetary disk was completed.
Otherwise, the protoplanet accumulates all gas around it before the gas dissipation when the gas disk exists for $\gg 10^5$\,yr.
In the latter case, the mass of the gas giant planet may be determined only by disk properties (i.e., the surface density and density profile of the protoplanetary disk).
Thus, to investigate the formation of the gas giant planet, we need to precisely estimate the properties of the protoplanetary disk, and the gas dissipation process (or timescale).

In this study, it was not possible to investigate gap formation in detail, since the simulations were limited in a local region around the protoplanet.
However, it was possible to resolve the present Jovian radius and the circumplanetary disk.
To understand gap formation and the mass accretion rate, it is necessary to simulate a protoplanetary system using a global simulation that can resolve the protoplanet.
However, such a calculation requires a large amount of CPU time. 
As well, issues concerning disk viscosity would need to be better understood before further investigation of both the `protoplanetary' and the `circumplanetary disk'  can be performed.
In general, it is considered that disk viscosity is caused by the magneto-rotational-instability (MRI). 
Thus, to properly calculate the viscous disk (circumplanetary and circumstellar disks), it is necessary to include magnetic effects.
Although it is currently possible to calculate the evolution of the magnetized disk with a lower spatial resolution  \citep{machida06b}, a simulation with a higher spatial resolution (or more CPU power) is necessary to investigate the disk viscosity correctly.
For such a simulation, the next generation of supercomputers is required.
However, the mass accretion rate derived in the present local simulation is consistent with those obtained in global simulations.
Thus, we could link the mass accretion rate in a global simulation to that in a local simulation. 
Therefore, we can safely estimate the growth of the gaseous protoplanet.

\section*{Acknowledgments}
We have greatly benefited from discussion with S. Watanabe, H. Tanaka, T. Tanigawa, and T. Muto. 
We are very grateful to an anonymous referee for a number of very useful suggestions and comments.
Numerical calculations were carried out using a Fujitsu VPP5000 at the
 Center for Computational Astrophysics, the National Astronomical
 Observatory of Japan. 
This work is supported by the Grants-in-Aid from MEXT (18740104, 20540238, 21740136).

\clearpage
\begin{table}   
\setlength{\tabcolsep}{3pt}
\caption{Model parameters}
\label{table:1}
\footnotesize
\begin{center}
\begin{tabular}{cccccccccccc}
\hline
Model & $\rht$   &  $\tl{r}_{\rm B}$ & $M_{\rm p}$  {\footnotesize (5.2 AU)} 
$^{a}$ & $\rho_{\rm cri}$ & $\tilde{r}_{\rm sink}$($10^{-3}$) &
\\
\hline
(M001A0, M001A1, M001A2, M001A3, M001I) & 0.29 &0.073 & 0.01  & (1, 10, $10^2$, $10^3$, $\infty$)& 3.5 \\
(M005A0, M005A1, M005A2, M005A3, M005I) & 0.5  &0.38 & 0.05  & (1, 10, $10^2$, $10^3$, $\infty$)& 3.5 \\
(M01A0, M01A1, M01A2, M01A3, M01I)     & 0.63 &0.75 & 0.1   & (1, 10, $10^2$, $10^3$, $\infty$)& 3.5 \\
(M02A0, M02A1, M02A2, M02A3, M02I)     & 0.8  &1.54 & 0.2   & (1, 10, $10^2$, $10^3$, $\infty$)& 3.5 \\
(M04A0, M04A1, M04A2, M04A3, M04I)     & 1.0  &3.0  & 0.4   & (1, 10, $10^2$, $10^3$, $\infty$)& 3.5 \\
(M06A0, M06A1, M06A2, M06A3, M06I)     & 1.15 &4.56 & 0.6   & (1, 10, $10^2$, $10^3$, $\infty$)& 3.5 \\
(M08A0, M08A1, M08A2, M08A3, M08I)     & 1.26 &6.0 & 0.8   & (1, 10, $10^2$, $10^3$, $\infty$)& 3.5 \\
(M1A0, M1A1, M1A2, M1A3, M1I)         & 1.36 &7.55 & 1     & (1, 10, $10^2$, $10^3$, $\infty$)& 3.5 \\
M02ISS       & 0.8 &1.54  & 0.2   & $\infty$ & 1.6 \\
M02ISM       & 0.8 &1.54  & 0.2   & $\infty$ & 2.5 \\
M02ISL       & 0.8 &1.54  & 0.2   & $\infty$ & 11 \\
\hline

\end{tabular}
\end{center}
$^{a}${in unit of Jupiter mass  $\mj$}
\end{table}

\clearpage
\begin{figure}
\begin{center}
\includegraphics[width=150mm]{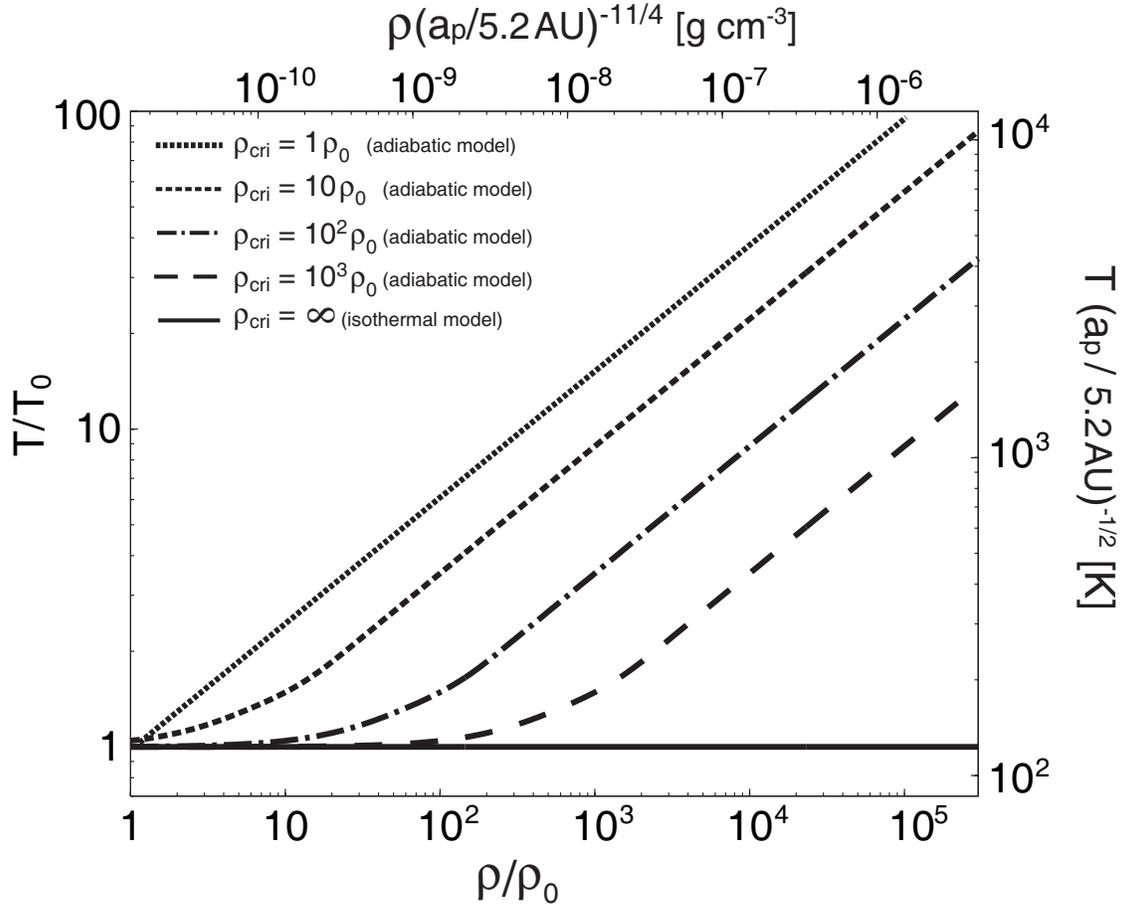}
\caption{
The density-temperature diagrams for the isothermal ($\rho_{\rm cri}=\infty$) and adiabatic ($\rho_{\rm cri}$ =$\rho_0$, 10$\rho_0$, $10^2\,\rho_0$, and $10^3\,\rho_0$) models.
The right and upper axes indicate the dimensional temperature and density at the Jovian orbit.
}
\label{fig:1}
\end{center}
\end{figure}

\clearpage
\begin{figure}
\begin{center}
\includegraphics[width=150mm]{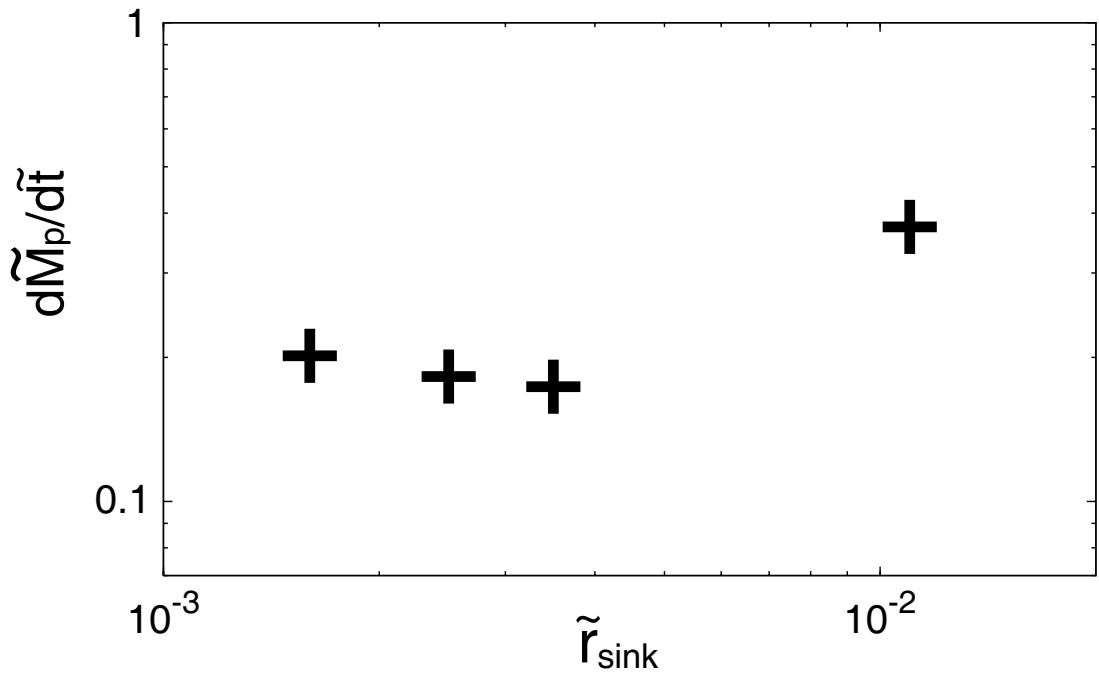}
\caption{
The mass accretion rate as a function of the sink radius for models M02ISS, M02ISM, M02ISL, and M02I.
}
\label{fig:2}
\end{center}
\end{figure}

\clearpage
\begin{figure}
\begin{center}
\includegraphics[width=150mm]{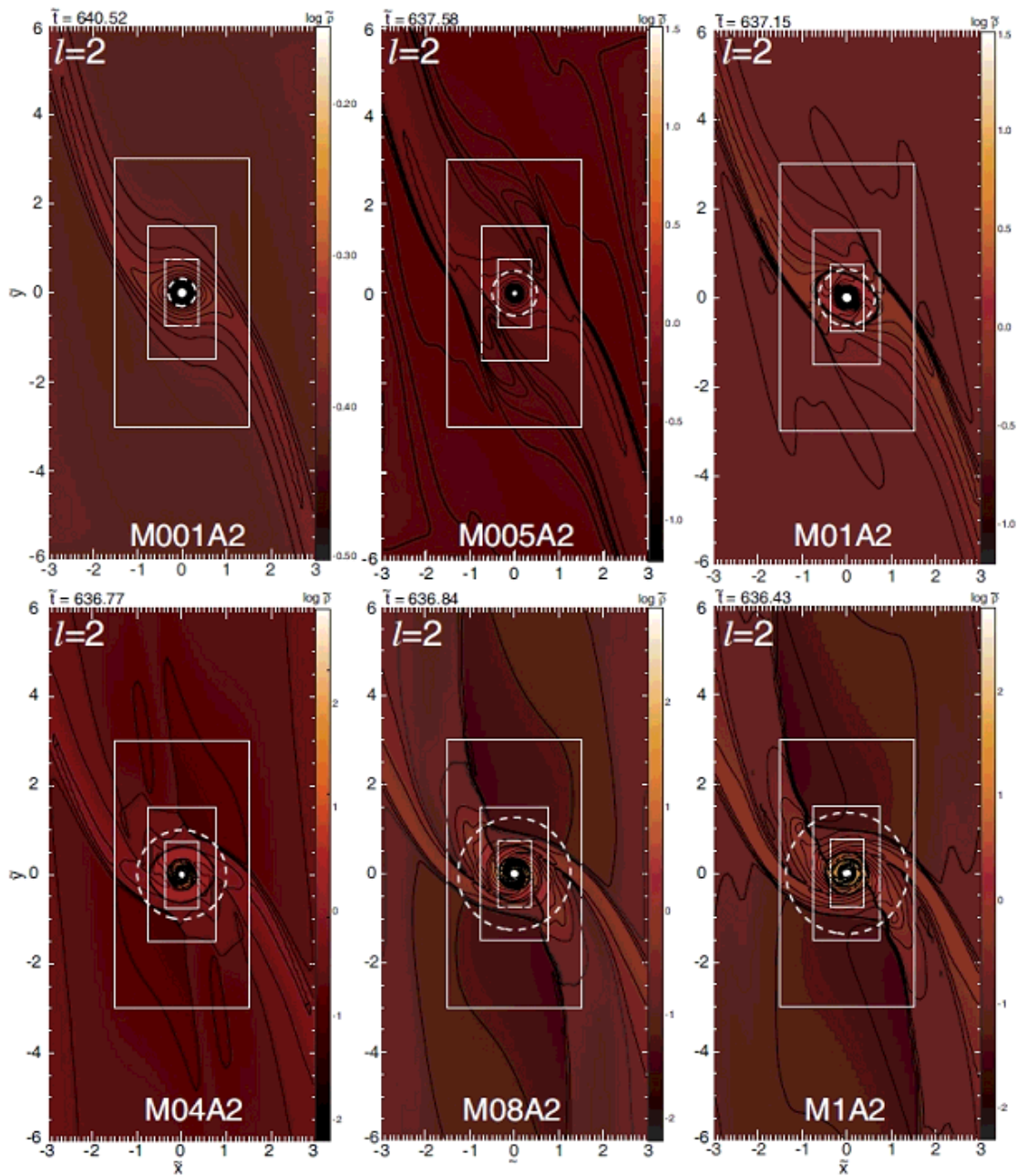}
\caption{
The density distribution ({\it color scale}) on the cross section in the $z=0$ plane for models M001A2, M005A2, M01A2, M04A2, M08A2, and M1A2.
The dashed circle represents the Hill radius.
Four grid levels are shown in each panel ($l$ = 2, 3, 4, and 5).
The level of the outermost grid is denoted in the top left corner of each panel.
The elapsed time $\tl{t}$ is denoted above each panel.
}
\label{fig:3}
\end{center}
\end{figure}

\clearpage
\begin{figure}
\begin{center}
\includegraphics[width=150mm]{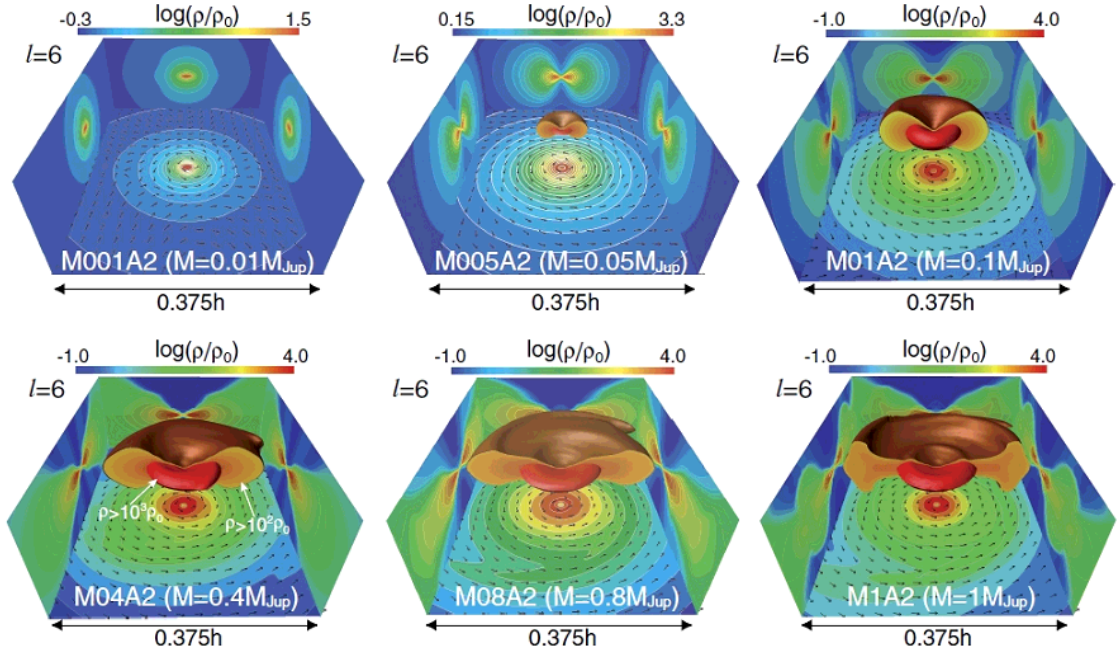}
\caption{
The gas structure around the protoplanet in a bird's-eye view for models M001A2, M005A2, M01A2, M01A2, M04A2, M08A2, and M1A2.
Density distribution on $x=0$, $y=0$, and $z=0$ plane are projected onto each wall surface.
The color surfaces indicate constant density surfaces: $\rho=10^{3}\,\rhoz$ (red) and $10^3\, \rhoz$ (orange).
Velocity vectors ({\it arrows}) are plotted on the bottom wall.
The size of the domains is shown in each panel.
The grid level is shown in the top left corner of each panel.
}
\label{fig:4}
\end{center}
\end{figure}

\clearpage
\begin{figure}
\begin{center}
\includegraphics[width=150mm]{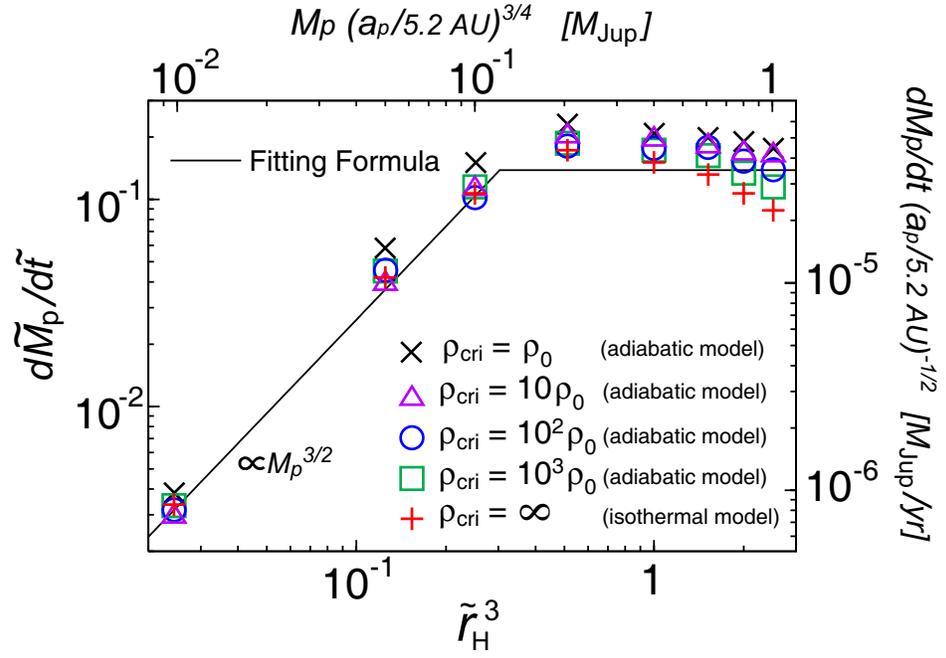}
\caption{
The mass accretion rate for models with $\rhoc$ = $\rhoz$($\times$), 10$\rhoz$ ($\triangle$), $10^2\,\rhoz$ ($\bigcirc$),  $10^3\,\rhoz$ ($\sq$), and  $\infty$ ($+$) as a function of the cubed Hill radius.
The right and upper axes indicate the dimensional mass accretion rate and the protoplanet mass at the Jovian orbit.
The solid line represents the fitting formula.
}
\label{fig:5}
\end{center}
\end{figure}

\clearpage
\begin{figure}
\begin{center}
\includegraphics[width=130mm]{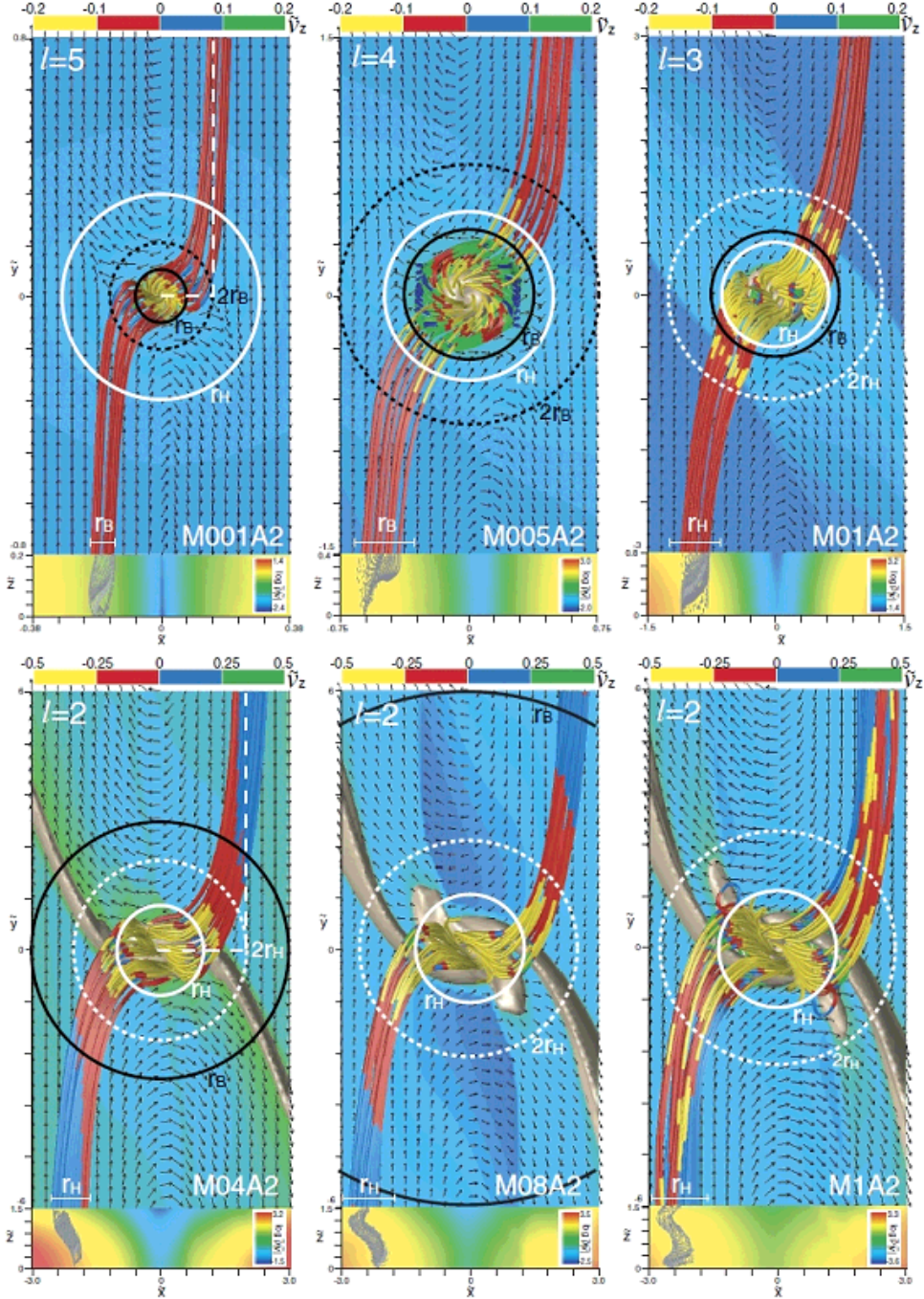}
\caption{
The gas streamlines and mass flux $\tl{\rho}\vert \tl{v} \vert$ for models M001A2, M005A2, M01A2, M04A2, M08A2, and M1A2.
{\it Each top panel}: The streamlines are plotted in three-dimensions.
The color of the streamlines indicates the vertical component of the velocity at each mesh point.
The values of the Bondi, doublewide Bondi, Hill and doublewide Hill sphere are plotted.
The grid level is shown in the top left corner of each panel.
{\it Each bottom panel}:
The mass flux in the $x$-$z$ plane is plotted on the bottom $y$-boundary.
The gas flows into the protoplanetary system are given only in the gray region.
}
\label{fig:6}
\end{center}
\end{figure}

\clearpage
\begin{figure}
\begin{center}
\includegraphics[width=150mm]{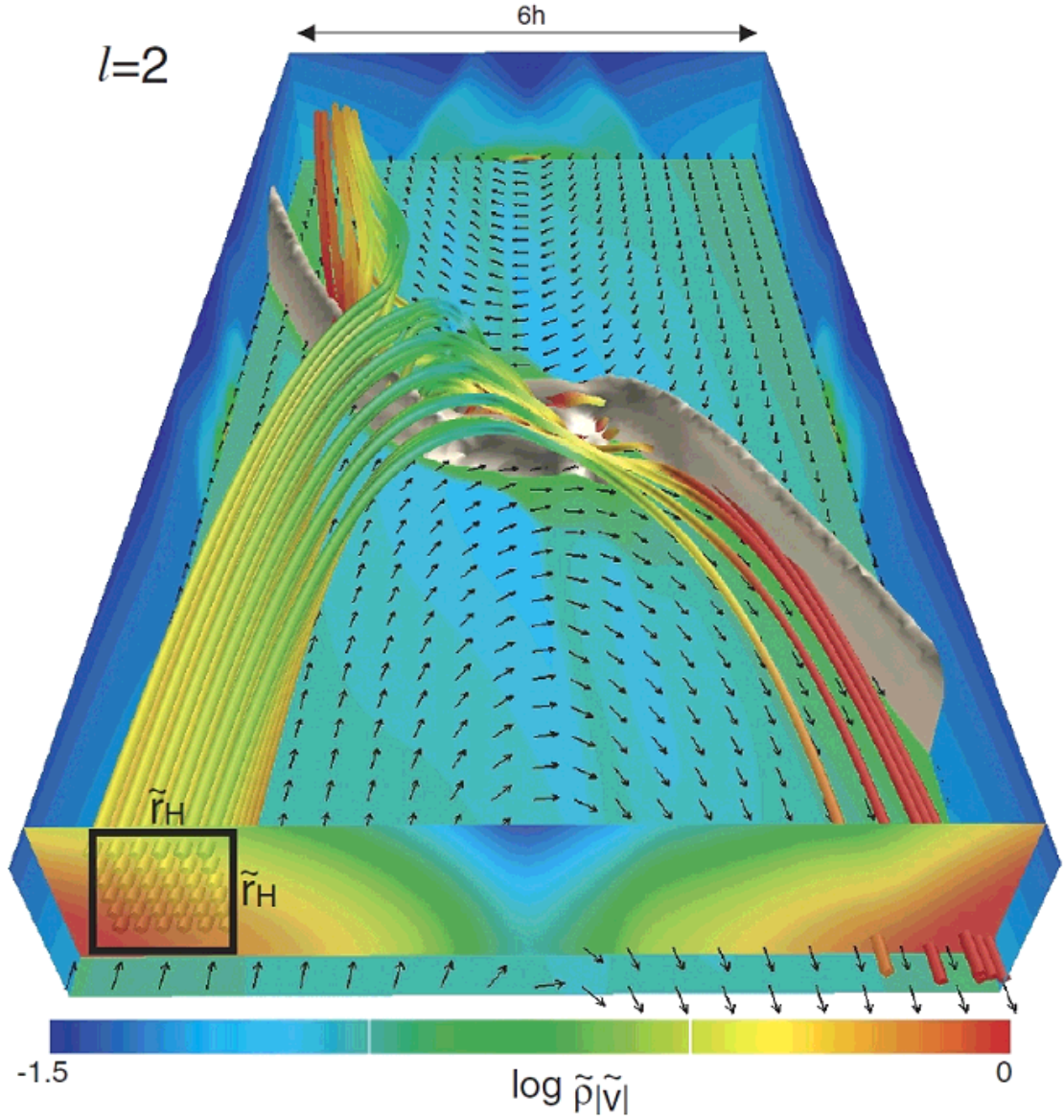}
\caption{
The gas streamlines integrated from the region inside the solid square for model M04A2.
The color of streamlines represents the mass flux at each mesh point.
Density distribution on the $x=0$, $y=0$, and $z=0$ planes are plotted on each wall surface.
The distribution of $\tl{\rho} \vert \tl{v} \vert$ on $y=-6\,h$ plane is also plotted.
The color surfaces indicate a constant density region of $\tl{\rho}=0.5$.
}
\label{fig:7}
\end{center}
\end{figure}

\clearpage
\begin{figure}
\begin{center}
\includegraphics[width=150mm]{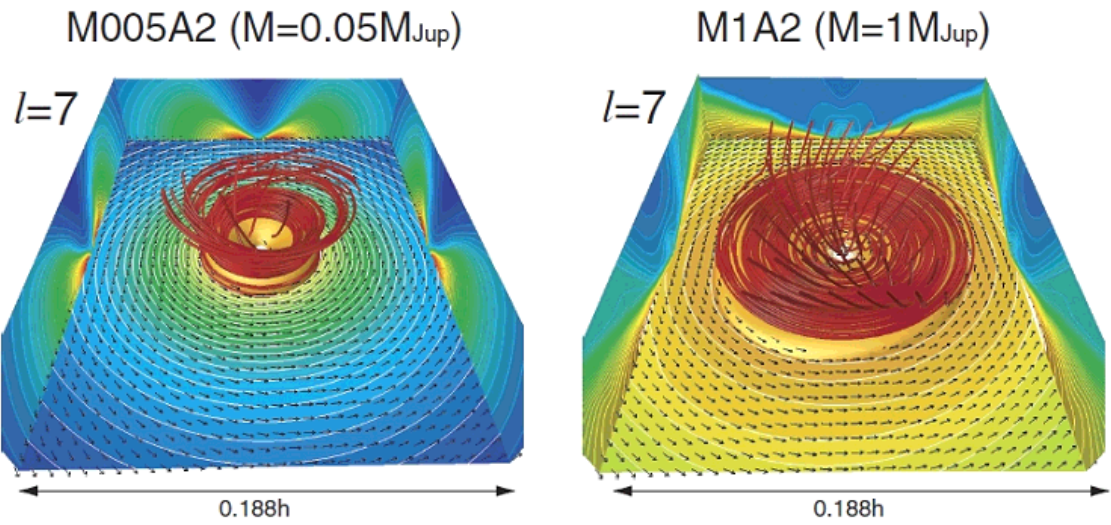}
\caption{
The gas structure around the protoplanet for models M005A2 and M1A2.
Density distribution on the $x=0$, $y=0$, and $z=0$ planes are projected on each wall surface.
The color surfaces indicate high-density surfaces.
The streamlines ({\it red lines}) are inversely integrated from the proximity of the protoplanet.
The size of the domains is shown in each panel.
The grid level is shown in the top left corner of each panel.
}
\label{fig:8}
\end{center}
\end{figure}

\clearpage
\begin{figure}
\begin{center}
\includegraphics[width=150mm]{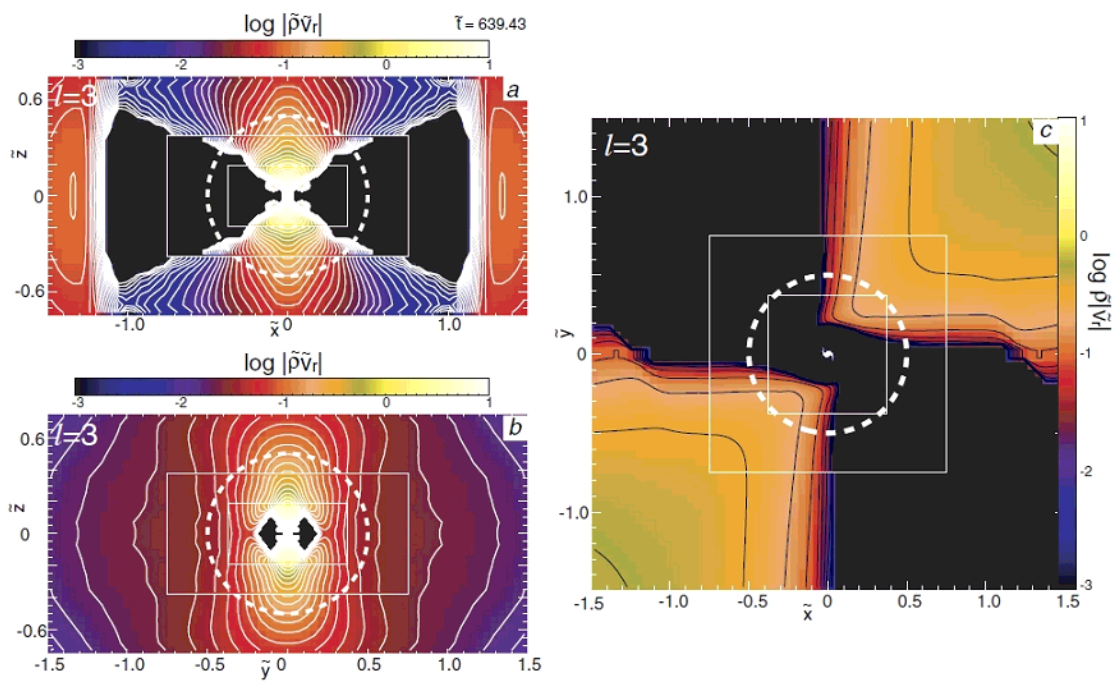}
\caption{
The mass flux ($\tl{\rho}\, \vert \tl{v}_r\vert$) around the protoplanet on the $y=0$ (top left), $x=0$ (bottom left) and $z=0$ (right) planes for model M005A2.
The gas in the colored region has a negative mass flux  $\tl{\rho}\,\vert \tl{v}_r \vert <0$, and flows into the protoplanetary system, while the gas in the black region has a positive mass flux $\tl{\rho}\, \vert \tl{v}_r\vert > 0$ and flows out from the protoplanetary system.
The dotted circle indicates the Hill radius.
}
\label{fig:9}
\end{center}
\end{figure}

\clearpage
\begin{figure}
\begin{center}
\includegraphics[width=150mm]{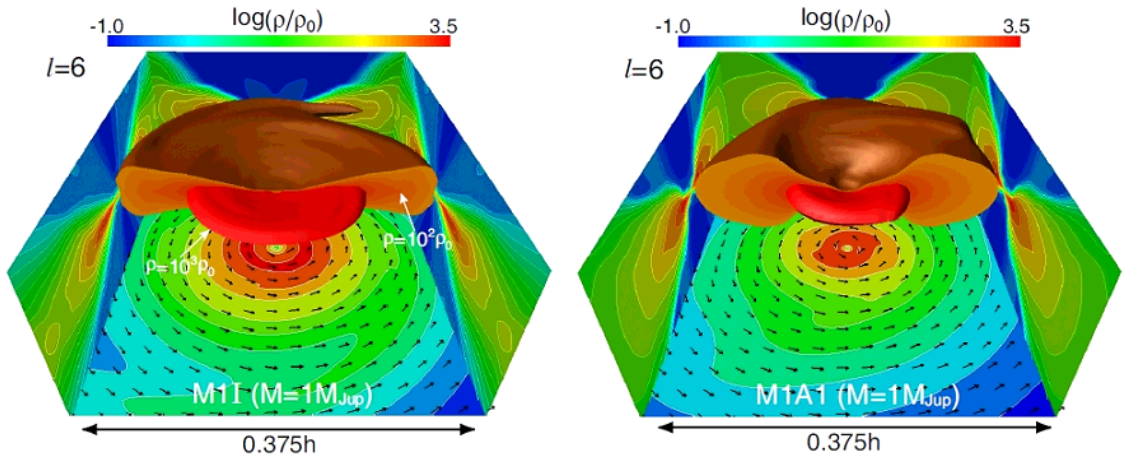}
\caption{
Same as Figure~\ref{fig:4}, but for model M1I (left) and M1A1 (right).
}
\label{fig:10}
\end{center}
\end{figure}

\clearpage
\begin{figure}
\begin{center}
\includegraphics[width=150mm]{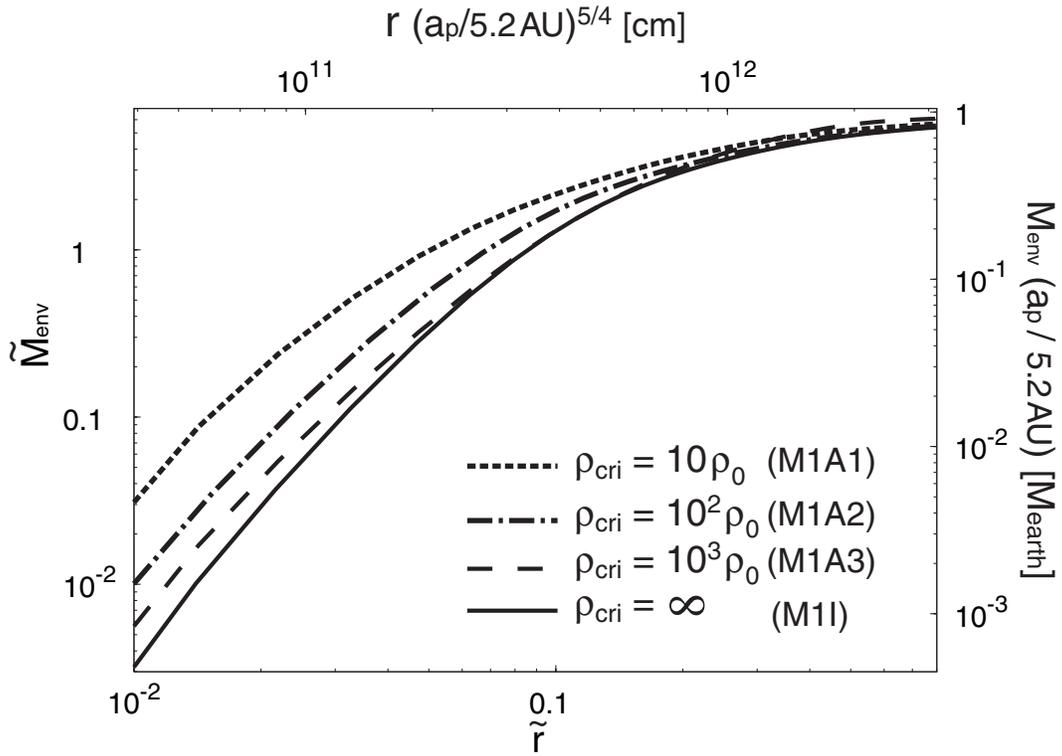}
\caption{
Cumulative mass integrated from the center for models M1A1, M1A2, M1A3, and M1I.
The right and upper axes indicate the dimensional mass and radius normalized by the Jovian orbit.
}
\label{fig:11}
\end{center}
\end{figure}

\clearpage
\begin{figure}
\begin{center}
\includegraphics[width=170mm]{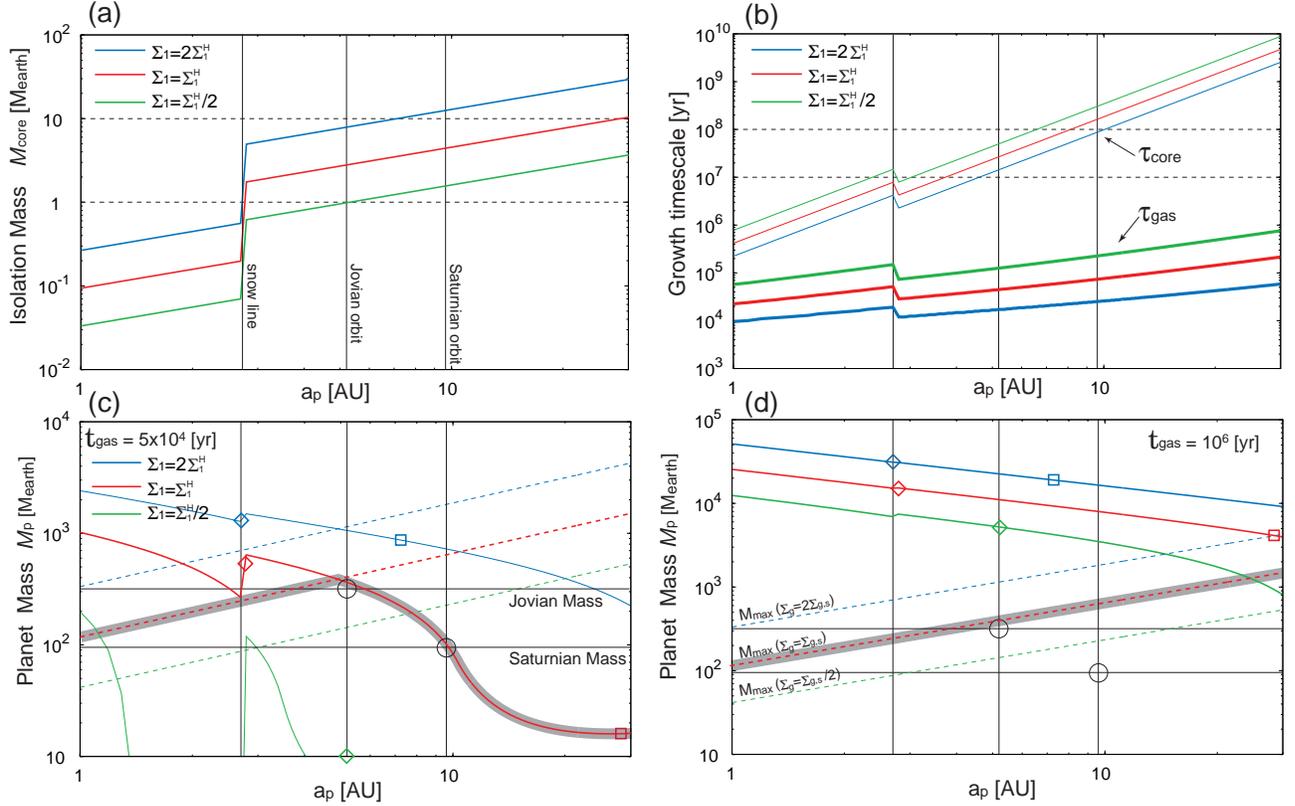}
\caption{
The isolation core mass, or mass of the planetary core, $M_{\rm core}$, ({\it b}) growth timescale for core $\tau_{\rm core}$ and Jovian-mass planet $\tau_{\rm gas}$, ({\it c}) Planet mass $M_{\rm p}$ when $t_{\rm neb}=5\times10^4$\,yr and (d) when $\tau_{\rm neb}=10^6$\,yr against the orbital radius $a_{\rm p}$. 
The dotted lines in panels {\it c} and {\it d} indicate the maximum gas mass acquired by the planet at each orbit.
The open diamonds and square in panels {\it c} and {\it d} indicate the orbital radius when the core mass exceeds $M>1M_\oplus$ (filled diamond) and $M>10M_\oplus$ (filled square).
Each open circle indicates the mass and orbital radius of Jupiter and Saturn.
}
\label{fig:12}
\end{center}
\end{figure}

\end{document}